\documentclass[preprint2]{aastex}
\usepackage{amsmath,amstext}
\usepackage{tikz,xspace}
\usepackage{fancyvrb}
\usetikzlibrary{shapes.geometric,arrows,positioning,calc}
\usepackage[breaklinks,colorlinks,citecolor=blue,linkcolor=magenta]{hyperref}
\usepackage[all]{hypcap} 

\newcommand{\msun}{M_{\odot}\xspace}
\newcommand{\mhsix}{M_{\rm h,6}\xspace}

\defcitealias{Li:2015a}{G. Li et al. 2015}
\defcitealias{Li:2015b}{S. Li et al. 2017}

\begin{document}

\shorttitle{A Self-Sustaining Black Hole Engine Powered by the Disruptions of Stars}

\shortauthors{Guillochon and Loeb}

\title{A Self-Sustaining Black Hole Engine Powered by the Tidal Disruptions of Stars}

\author{James Guillochon\altaffilmark{1,2} and Abraham Loeb\altaffilmark{1,3}}
\altaffiltext{1}{Harvard-Smithsonian Center for Astrophysics, The Institute for Theory and
Computation, 60 Garden Street, Cambridge, MA 02138, USA}
\altaffiltext{2}{ORCID: \url{https://orcid.org/0000-0002-9809-8215}}
\altaffiltext{3}{ORCID: \url{https://orcid.org/0000-0003-4330-287X}}

\email{guillochon@gmail.com}

\begin{abstract}
Tidal disruption events (TDEs) strongly prefer host galaxies undergoing, or recovering from, a burst of star formation, implying that a rare minority of galaxy types produces most events. The required per-galaxy rates, as high as $10^{-2}$ gal$^{-1}$ yr$^{-1}$, are hard to achieve through stellar relaxation alone. Molecular clouds surrounding nuclear clusters help set the relaxation rate, offering a path to higher rates. We show that (post-)starburst galaxies, whose nearby prototypes contain large molecular gas reservoirs, are likely in a self-sustaining cycle: an enhanced disruption rate compresses the surrounding clouds through momentum injected by the unbound debris, the denser clouds compress the cluster, and the cluster disrupts stars faster still. The runaway is arrested only when stars begin to collide. Solving for the steady state, the rate saturates at $1.4\times10^{-2}(M_{\rm h}/10^{6}M_\odot)^{-0.84}$ yr$^{-1}$, two orders of magnitude above the canonical rate at $10^{6}M_\odot$, with a normalization uncertain by a further two orders of magnitude through the stellar collision rate. Three independent requirements --- that tidal debris outweigh AGN feedback, that the molecular clouds fit within the disk that holds them, and that the cusp be no denser than observed nuclei --- bound the flattening of the stellar cusp to $0.18 \lesssim f_\ast \lesssim 0.25$. Because no engine can run below a threshold black hole mass, holes seeded beneath it stay dark until accretion carries them across, switching on after $\sim 1$ Gyr and offering a natural explanation for the late peak recently measured in the TDE delay time distribution. The nucleus is buried under $A_V \simeq 50$ whatever its geometry, so most such disruptions should be hidden from optical surveys and emerge instead in the infrared.
\end{abstract}


\section{Introduction}\label{sec:intro}

Black holes grow by feeding on gas delivered to them from their surroundings; this gas is either delivered to them in the form of low-density flows that funnel through an extended accretion disk structure \citep{Salpeter:1964a}, or by the disruptions of gravitationally self-bound structures such as stars and planets \citep{Rees:1988a}. In both cases, the black hole emanates light and mass (in the form of outflows) which can feed energy and momentum into its neighborhood \citep{Fabian:2012a, Metzger:2015a, Jiang:2016b}. The majority of the energy released by an accreting black hole is in the form of light unless the light is trapped within the accretion flow, which occurs when the accretion rate exceeds the Eddington limit $L_{\rm Edd} \equiv 4 \pi G M_{\rm h} m_{\rm p} c / \sigma_{\rm T}$; above this limit, jets and outflows carry away similar amounts of energy.

\begin{figure*}[t]
\centering\includegraphics[width=0.8\linewidth,clip=true]{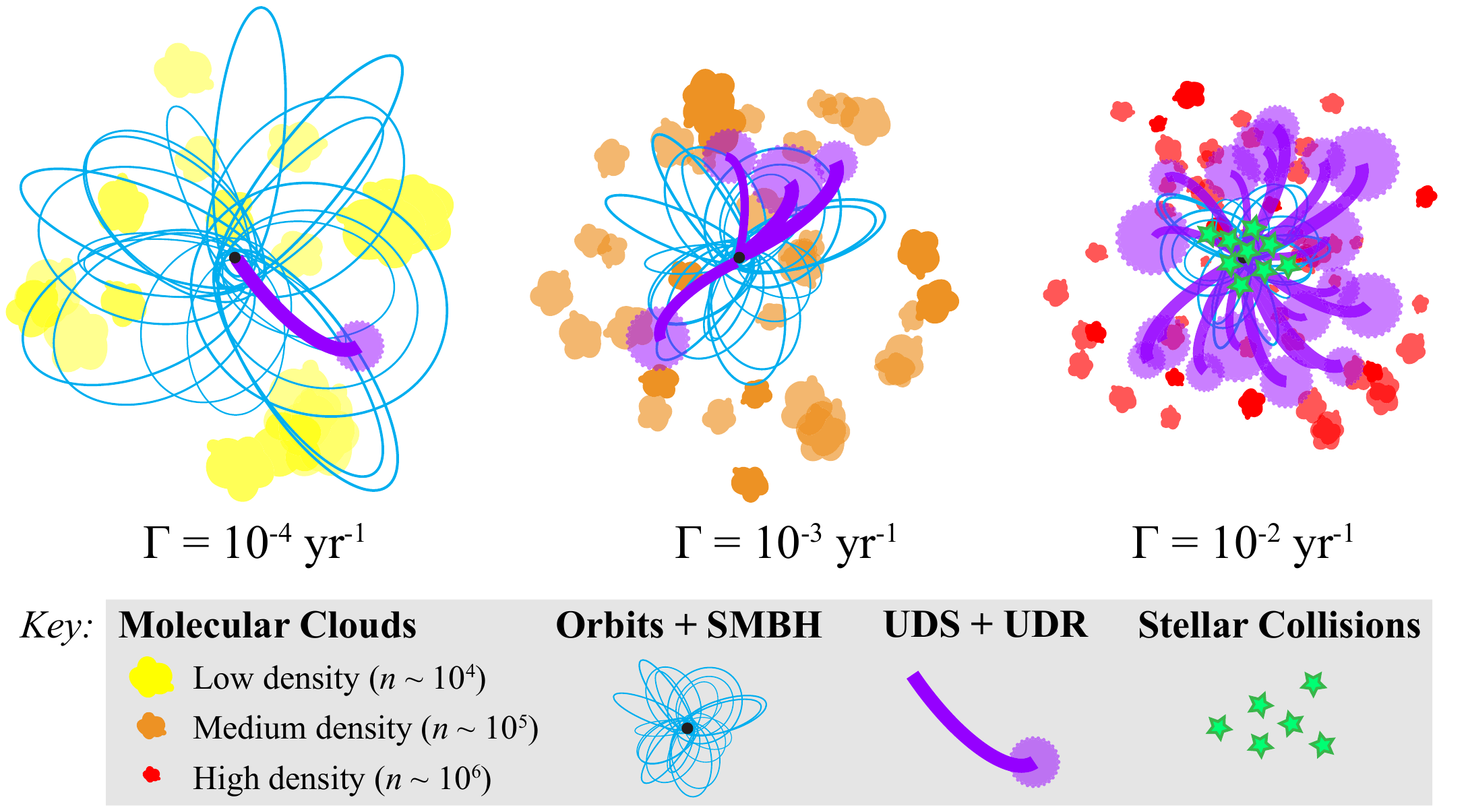}
\caption{Evolution of the nuclear cluster and surrounding dense molecular clouds as the rate of disruptions increases, with the left panel showing the conditions arising from stellar relaxation alone, the middle panel showing a nuclear cluster first entering the engine state, and the right panel showing the equilibrium conditions established once stellar collisions inhibit the rate of tidal disruption. The key shown at the bottom of the illustration highlights the components featured in each panel. From left to right, both the cloud density and stellar density of the nuclear cluster increase with an increasing tidal disruption rate.}
\label{fig:diagram}
\end{figure*}

For driving turbulence however, the primary resource required is momentum, and light carries far less momentum than outflowing matter moving at velocity $v$ by a factor $v/c$. The energy carried by light can be converted into momentum, but only if it is able to couple to the gas and does not escape to infinity first. Because it is not bound to anything but relativistic gravitational potential wells, light is inherently difficult to trap completely, and will reorganize the structure of the surrounding matter, often times blowing it out of the system entirely, in order to escape. Fast-moving flows (e.g. jets) suffer a similar problem: while they may carry a lot of kinetic energy per baryon, they can punch through surrounding matter with ease and escape before depositing momentum into their surroundings.

The best way to deliver ample amounts of momentum is within a massive flow of modest velocity, such flows will have difficulty escaping shallow potential wells before being able to deposit their momenta. This is why a single supernova, which only converts a fraction of a percent of its rest mass into kinetic energy, can be more effective at driving turbulence than a $10^{6} M_{\odot}$ black hole (which converts $\sim$ 10\% of incoming mass into energy) accreting at its Eddington limit for hundreds of years, as the typical velocity of the former (thousands of km~s$^{-1}$) is significantly smaller than the latter ($\sim c$).

The disruptions of gravitationally self-bound objects are almost always accompanied by the ejection of $\sim 50$\% of the disrupted object's mass. For the case of a star being disrupted by a supermassive black hole, the kinetic energy carried by this effluence is comparable to the energy of the ejecta associated with the explosion of a massive star \citep[on average a factor of $\sim 5$ times smaller,][]{Guillochon:2016b}. When this outflow interacts with the ambient gas, it deposits its energy into an expanding bubble that mimics a supernova remnant, these unbound debris remnants (UDR) should be observable in our own galactic center and the centers of nearby galaxies \citep{Khokhlov:1996a,Guillochon:2016b,Chen:2016b,Romero-Canizales:2016a}.

In this paper, we show that the momentum deposited by UDRs can exceed the momentum deposited by both supernovae (SNe) and active galactic nuclei (AGN) when the stellar disruption rate exceeds a critical value. When these conditions are met, we argue that a steady-state condition is established that results in a sustained enhancement of the tidal disruption rate by a factor of hundreds, lasting for as long as the nucleus retains its fuel, which we show in Section~\ref{sec:discussion} to be $\sim 10^{8}$~yr. This enhancement in the disruption rate arises from the increase in density of the molecular clouds that surround the black hole which is driven by the disruptions themselves, which act to compress the underlying nuclear cluster of stars (Figure \ref{fig:diagram}).

When a black hole is in this state, we argue that the host galaxy will exhibit a number of identifiable features that have already or are expected to be revealed in observations of tidal disruption host galaxies:
\begin{itemize}
\item Because the state requires the disruption rate to be enhanced above a critical value, a violent relaxation process likely needs to have occurred to ``spark'' the engine state. A supermassive black hole merger is an example of such a process, and if it is responsible for sparking the engine the galaxy should show signs of a recent merger and its associated starburst. Such post-starburst galaxies have already been identified observationally as being strongly tied with tidal disruptions \citep{Arcavi:2014a,French:2016a}, an over-representation by roughly two orders of magnitude relative to their share of the galaxy population that has survived a decade of subsequent survey work \citep{French:2020a,Gezari:2021a,Hammerstein:2023a}. The age dependence of that enhancement is now measured: \citet{Shepherd:2026a} construct the delay time distribution from 41 TDE hosts and find the rate rising with post-burst age to a peak near $1$~Gyr. This is a demanding constraint on any sparking mechanism, and most models fail it by predicting a rate that declines from the burst onwards. The engine does not: because it requires a minimum black hole mass, a hole seeded below that threshold switches on only after it has grown into the engine regime, which introduces a delay of the observed order. We develop this in Section~\ref{subsec:lifetime}.
\item The black hole will be surrounded by molecular clouds with densities that are much higher than usual, and that dense molecular gas tracers should reveal the presence of these clouds. Post-starburst galaxies routinely show such central concentrations of molecular gas \citep{French:2015a,Smercina:2018a}, with NGC~1266 the best-studied local example \citep{Alatalo:2011a,Alatalo:2014a}.
\item Both the stellar density and stellar velocity dispersions in the nucleus should be enhanced relative to a galaxy that is not in the engine state. Such an enhancement has been claimed to be detected in one of the nearest k+A galaxies \citep{Stone:2016b}.
\item The surrounding SNe rate will be suppressed relative to the expected rate given a gas surface density as only the very-densest clouds will collapse to form stars. Suppression of the star formation rate has been observed in the centers of post-starburst galaxies \citep{French:2018a}, and in NGC~1266 in particular \citep{Alatalo:2015a}.
\item Because the black hole will be surrounded by a significant gas reservoir, it will accrete some fraction of this gas and is likely to be a low-luminosity AGN (LLAGN), with the resulting accretion rate being larger than the late-time fallback accretion produced by the disruptions. The host of the TDE ASASSN-14li has pre-flare radio emission consistent with being a LLAGN \citep{Alexander:2016a,Alexander:2020a}.
\end{itemize}

In Section~\ref{sec:sources} we quantify and compare the amount of feedback from SNe, AGN and TDEs. In Section~\ref{sec:engine} we describe the self-sustaining black hole engine state and the conditions required to enter it, followed by Section~\ref{sec:equil} where we calculate the equilibrium tidal disruption rate expected within this state. In Section~\ref{sec:spark} we describe potential ways of temporarily enhancing the disruption rate in order to satisfy the conditions set in Section~\ref{sec:engine}. We summarize the observational characteristics of TDE engines and their influence on black hole and galaxy evolution in \ref{sec:discussion}. Many quantities presented in this work are written in normalized form, which is indicated by a numerical subscript, e.g. $\mhsix$ is the black hole mass divided by $10^{6}$ solar masses.

\section{Sources of driving in galactic centers}\label{sec:sources}

In the absence of tidal disruptions, the two primary actors that drive turbulence in the clouds surrounding a galactic nucleus are SNe and AGN activity. While some SNe originate from long-lived channels that don't depend much on the current rate of star formation, (e.g. type Ia SNe), the SNe that are most likely to be within star-forming regions are the massive stars associated with ongoing star formation. As a result of this connection, SNe are intimately related with star formation, which makes them a natural limiter on the star formation rate in a molecular cloud; too many SNe, and a cloud will be blown apart before being able to form more stars. SNe are critical in setting the balance between gravitational collapse and uncontrolled expansion, but this balance is disturbed near a supermassive black hole, as an active black hole can easily release a supernova's worth of energy accreting at its Eddington limit for about a year. However, most black holes are not active, especially in the local universe, and even at the peak of AGN activity at $z \sim 2$ the quasar fraction (i.e. massive black holes accreting at their Eddington limits) did not exceed 10\% \citep{Martini:2013a}. As a result, supernovae are a dominant force for driving turbulence in most galaxies.

Tidal disruptions change this picture further: now even inactive black holes can disturb their surroundings upon destroying infalling stars rather than relying on the continual accretion of nearby gas. Tidal disruptions are not common under steady-state conditions and canonically occur once per $10^{4}~{\rm yr~gal}^{-1}$, with even the most favorable rates not exceeding $\sim 10^{-3}$~yr$^{-1}$ \citep{Magorrian:1999a,Wang:2004a,Stone:2016a,Stone:2020a}, a rate comparable to the rate of SNe in the central kpc of our own galaxy. Volumetric rates measured by wide-field optical surveys are an order of magnitude below these loss-cone predictions \citep{vanVelzen:2021a,Hammerstein:2023a,Yao:2023a}, a discrepancy that is at least partly explained by dust obscuration of events in gas-rich nuclei \citep{Masterson:2024a}, a point we return to in Section~\ref{sec:discussion}. However, there are a number of potential ways of enhancing the tidal disruption rate temporarily, including anisotropy of the nuclear cluster \citep{Merritt:2004a,Lezhnin:2016a}, and the mergers of two black holes (\citealt{Chen:2011a}; \citetalias{Li:2015a}; \citetalias{Li:2015b}). We will discuss how the tidal disruption rate might be coaxed to such high values in Section~\ref{sec:spark}.

If the tidal disruption rate could be enhanced arbitrarily, the feedback from the outgoing debris and accretion resulting from stellar tidal disruptions could become more important than both SNe and AGN activity combined. Because tidal disruptions inject an energy comparable to a single supernova, which dominate feedback in most galactic nuclei which are inactive, the rate required for TDEs to dominate all other feedback processes is not much beyond the canonical $10^{-4}$~yr$^{-1}$ rate typically quoted for tidal disruptions.

To determine what rates are required for TDEs to dominate feedback relative to both SNe and AGN activity, we consider the amount of momentum injected into the surrounding ISM by all three processes. Generically, the net momentum (summed vectorially) deposited by a perfectly symmetrical blastwave in a static background is exactly zero, and thus the appropriate conserved quantity to consider is the kinetic energy deposited by the blast. The blastwave will sweep up the ISM in its wake, sharing its kinetic energy with a total mass that is excess of the ejecta mass by a factor of several, resulting in a momentum deposition that can exceed $M_{\rm ej} v_{\rm ej}$ by a factor $M_{\rm ISM}/M_{\rm ej}$, where $M_{\rm ej}$ is the ejecta mass, $M_{\rm ISM}$ is the swept-up mass, and $v_{\rm exp}$ is the velocity of the explosion at $t = 0$. However, this energy is only useful for driving turbulence if it can be converted into turbulent kinetic energy before leaking as radiation to infinity. This means that the total momentum injection per explosion scales as $M_{\rm ej} v_{\rm ej}^{2}/ v_{\rm rad}$, where $v_{\rm rad}$ is the velocity of the expanding shell at the time the it becomes radiative, which is a factor $v_{\rm ej}/ v_{\rm rad}$ times larger than the initial momentum of the blast. This generic result applies to all forms of kinetic energy injection, whether it be from a supernova, a jet, or the tail of unbound debris resulting from a tidal disruption.

\subsection{Supernovae}\label{subsec:sne}

Supernovae can have a wide range of energies, but the majority of observed supernovae have energies $\sim 10^{51}$~ergs, a small fraction of their rest mass. In star-forming molecular clouds, momentum injection is dominated by core-collapse SNe (CCSNe), the rate of which is proportional to the star formation rate and thus the surface gas density quantified by the Kennicutt-Schmidt law \citep{Kennicutt:1998a},
\begin{equation}
\Sigma_{\rm SFR} \simeq 2.5 \times 10^{-4} \left(\frac{\Sigma_{\rm gas}}{M_{\odot} {\rm pc}^{-2}}\right)^{1.4}~M_{\odot}~{\rm yr}^{-1}~{\rm kpc}^{-2}.
\end{equation}

If we assume a Kroupa IMF \citep{Kroupa:2001a} and that all stars formed with masses greater than 8~$M_{\odot}$ will produce a core-collapse SNe, roughly one-third of the star-forming mass will go into CCSNe. With most of the supernovae coming from stars of mass $\sim 10 M_{\odot}$, this yields a surface CCSNe rate of
\begin{equation}
\sigma_{\rm CCSNe} \sim 10^{-6} \left(\frac{\Sigma_{\rm gas}}{M_{\odot} {\rm pc}^{-2}}\right)^{1.4}~{\rm yr}^{-1}~{\rm kpc}^{-2}\label{eq:gammasne}
\end{equation}

The surface density of gas in the Milky Way interior to 10~kpc is $\sim 10 M_{\odot}$~pc$^{-2}$ \citep{Kalberla:2009a}, and thus from the above expression we expect that the GC should be producing CCSNe at a rate $\Gamma_{\rm CCSNe} = 10^{-4}$ per year in the central kpc. Assuming that $3 \times 10^{5} M_{\odot}$~{\rm km~s}$^{-1}$ of momentum is injected per SN \citep{Kim:2015a}, this results in a momentum injection rate of 

\begin{align}
\dot{p}_{\rm CCSNe} &= 2 \times 10^{32} \left(\frac{\Gamma_{\rm CCSNe} (r < {\rm kpc})}{10^{-4}~{\rm yr}^{-1}}\right)~{\rm g~cm~s}^{-2}\label{eq:psne}.
\end{align}
In a starbursting galaxy, the star formation rate can be a hundred times greater per unit area than the Milky Way, meaning that a momentum injection rate of $\dot{p}_{\rm CCSNe} \simeq 10^{34}$~{\rm g~cm~s}$^{-2}$ is possible in the most extreme cases.

Our use of momentum rather than energy as the means of feedback deserves an explanation, as it applies equally to the UDRs of Section \ref{subsec:tde}. A remnant expanding into gas of the densities we derive below, $10^{4}$--$10^{5}~{\rm cm}^{-3}$, has a post-shock cooling time of only a few years, four to five orders of magnitude shorter than the crossing time of the clouds it is driving. Roughly $90\%$ of the initial energy is radiated away before the remnant merges with its surroundings \citep{Cioffi:1988a,Blondin:1998a,Thornton:1998a}, and the remnant passes quickly from the adiabatic phase into a momentum-conserving snowplow. What survives is the terminal momentum, which is insensitive both to the ambient density, scaling only as $n^{-0.17}$, and to the clumpiness of the medium, which alters it by less than $60\%$ \citep{Kim:2015a,Martizzi:2015a,Walch:2015a}. Because the hot phase cools within years and occupies a fraction $\sim 10^{-6}$ of the gas volume at any instant, thermal pressure cannot contribute appreciably to the support of the clouds, and momentum deposition is the relevant channel. We note that this conclusion is specific to isolated remnants expanding into dense gas; clustered explosions in a pre-excavated cavity retain considerably more of their energy and can remain over-pressured for much longer \citep{Gentry:2017a}.

This argument is what licenses the energy balance we impose in Section \ref{subsec:conditions}, where the injected UDR energy is radiated by the cold, CO-emitting phase rather than retained as bubble pressure. The two statements are consistent rather than double-counted: the short cooling time is precisely what transfers the energy out of the hot shocked gas and into the $10$~K molecular phase, where it must then be radiated away for the cloud to remain in a steady state.

\subsection{Active Galactic Nuclei}\label{subsec:agn}

AGN can feed back into their surroundings via either radiation or outflows, with outflows being subdivided into ``winds'' and highly-collimated jets. Most of the energy expelled by the black hole comes out as light with typical efficiencies $\epsilon \sim 10\%$, but this light may not fully couple with the surrounding gas before being able to leak to infinity. And because light moves at such a high velocity, the momentum that light carries is small per unit energy. For photons, momentum deposition critical depends on the optical depth $\tau$, which is typically order tens for AGN \citep{Nenkova:2008b} and scales to first order proportionally to $\dot{M}$ \citep{Netzer:2006a},
\begin{align}
L_{\rm rad} &= 1 \times 10^{44} \lceil f_{\rm Edd} \rceil \epsilon_{d,-1} \mhsix~{\rm ergs~s}^{-1}\\
\tau &= \kappa_{\rm d} n_{\rm MC} m_{\rm p} a_{\rm h} \simeq 30 f_{\rm Edd}\label{eq:agntau}\\
\dot{p}_{\rm rad} &= \frac{\tau L_{\rm rad}}{c}\nonumber\\
&=1 \times 10^{35} \lceil f_{\rm Edd} \rceil f_{\rm Edd} \epsilon_{d,-1} \mhsix ~{\rm g~cm~s}^{-2}\label{eq:pagnrad}
\end{align}
where $f_{\rm Edd} \equiv L/L_{\rm Edd} = M/M_{\rm Edd}$ and $\lceil f_{\rm Edd} \rceil \equiv \min(f_{\rm Edd},1)$, and we have assumed that the nuclear cluster size $a_{\rm h} \sim 1~{\rm pc}$, an effective dust opacity $\kappa_{\rm d} \sim 100$ \citep{Li:2001a}, and typical molecular cloud number density $n_{\rm MC} = 10^{4}$.

Jets can also mechanically inject momentum into the surrounding gas,
\begin{align}
L_{\rm jet} &= 1 \times 10^{44} f_{\rm Edd} \epsilon_{j,-1} \mhsix~{\rm ergs~s}^{-1}\\
\dot{p}_{\rm jet} &= \frac{L_{\rm jet}}{c}\nonumber\\
&=3 \times 10^{33} f_{\rm Edd} \epsilon_{j,-1} \mhsix~{\rm g~cm~s}^{-2}\label{eq:pagnjet}.
\end{align}
Jets have an advantage that their luminosity is not restricted to be below Eddington, and that their coupling with the gas scales more weakly with $f_{\rm Edd}$ as $f_{\rm Edd} \rightarrow 0$. However, if a jet's direction remains unchanged over long periods of time, it can clear a channel of gas alone the jet direction that can prevent efficient coupling \citep{Heinz:2014a}. 

Nearly-isotropic winds released by the disk can couple more efficiently with the surrounding gas \citep{King:2015a}, and have the additional advantage of carrying more momentum per unit energy. In fact, winds are so effective that if their efficiency constant was much larger than the fiducial value of $\epsilon_{\rm w} = 10^{-4}$, they would easily expel most of the gas from the nucleus \citep{Ostriker:2010a}; with this assumption the energy and momentum injected by the wind is
\begin{align}
L_{\rm wind} &= 10^{41} f_{\rm Edd} \epsilon_{w,-4} \mhsix~{\rm ergs~s}^{-1}\\
\dot{p}_{\rm wind} &= 10^{32} f_{\rm Edd} \epsilon_{w,-4} \mhsix~{\rm g~cm~s}^{-2}\label{eq:pagnwind}.
\end{align}

Comparison of Equations (\ref{eq:pagnrad}), (\ref{eq:pagnjet}), and (\ref{eq:pagnwind}) to Equation (\ref{eq:psne}) shows that if even a black hole is slightly active with $f_{\rm Edd} = 10^{-2}$, the AGN will dominate supernova feedback even in a starbursting galaxy.

\subsection{Stellar Collisions}\label{subsec:coll}

Once a nuclear cluster is compressed to the densities we consider below, physical collisions between stars become frequent. Collisions are not an important source of momentum for the surrounding gas: a grazing encounter at the velocities of interest unbinds at most a few percent of the stellar envelope \citep{Lai:1993a,Freitag:2005a}, so that the momentum released per collision is smaller than that of a UDR by three orders of magnitude. Their importance is instead dynamical. A collision randomizes the orbital angular momentum of the participating stars (or merges them), removing them from the loss cone before they can be disrupted. Stellar collisions therefore act as the negative feedback that closes the engine loop, and we return to them quantitatively in Section~\ref{sec:collcond}.

\subsection{Tidal Disruptions}\label{subsec:tde}
As described in Section \ref{sec:sources}, the canonical tidal disruption rate predicted for galactic nuclei is $10^{-4}~{\rm yr}^{-1}$. However, if the hypothesis that tidal disruptions originate from a small fraction of all galaxies, the tidal disruption rates in those galaxies is significantly larger, and therefore we normalize the disruption rate $\Gamma_{\rm TDE}$ to $10^{-2}~{\rm yr}^{-1}$, $\Gamma_{\rm TDE, -2}$, in the following section. Even for recently starbursting nuclear environments, most stars disrupted by a SMBH will be low-mass stars \citep{Kochanek:2016a}, and in the following section we assume an average star mass of 0.5 $M_{\odot}$. As most disrupted stars will originate from the SMBH's sphere of influence \citep{Lightman:1977a}, they approach the black hole on parabolic orbits with orbital energies that are a factor $a_{\rm h} r_{\rm t}^{2} / R_{\ast} \sim 10^{-4}$ the energy spread across the stellar debris, and thus almost exactly half of the star mass accretes onto black hole, and half is ejected within an unbound debris stream.

It has been suggested previously that the unbound debris can inject tremendous amounts of energy into the gas that surrounds SMBHs \citep{Khokhlov:1996a}. \citeauthor{Guillochon:2016b} showed that the amount of energy injected by a typical disruption is comparable to that of a supernova, approximately $10^{50}~{\rm ergs}$ per disruption. As supernovae are important drivers of turbulence in molecular clouds, the similarity in energy means that tidal disruptions should be seriously considered as a sources of feedback in galaxies, especially in their nuclei. This has recently been enforced by observational evidence from the TDE-hosting-galaxy ASASSN-14li that the unbound debris may be responsible for producing off-nuclear radio emission at pc scales \citep{Romero-Canizales:2016a}.

As an unbound debris stream (UDS) interacts with the ambient medium, it produces a remnant similar to that of a supernova, an unbound debris remnant (UDR), in which the kinetic energy of the UDS has been converted into work done on the surrounding ISM and into heat. Because the tidal disruption radius grows slowly relative to the Schwarzschild radius, stars move faster when they are disrupted by more-massive black holes, with the speed approaching $c$ for events where $r_{\rm t} \sim r_{\rm g}$. As a result, the kinetic energy of the unbound debris has a weak scaling with the SMBH mass $\propto M_{\rm h}^{1/3}$, which translates to more energy being contained in the resulting UDR, meaning that more massive black holes yield more energy per disruption than less massive ones. If disruptions occur with regular frequency, this translates into an energy and momentum deposition rate of
\begin{align}
L_{\rm UDR}  &= 3 \times 10^{40} \Gamma_{-2} \mhsix^{1/3}~{\rm ergs~s}^{-1}\\
\dot{p}_{\rm UDR}  &= 1 \times 10^{33} \Gamma_{-2} \mhsix^{1/3}~{\rm g~cm~s}^{-2}\label{eq:pudr},
\end{align}
where just as in the case of SNRs, the final momentum per UDR injected into the ISM is dependent upon the velocity of the flow at the time the remnant becomes radiative.
 
As the black hole accretes matter from the bound half of the debris, it will radiate away the accretion energy, which can also feed back on the environment. Tidal disruptions can be quite inefficient at converting mass into radiation simply because they can exceed the Eddington limit at peak by a few orders of magnitude, although this becomes less problematic for black holes in excess of $10^{7}~M_{\odot}$. Taking $t_{\rm peak} = 0.1~M_{\rm h,6}^{1/2}$ \citep{Guillochon:2013a} and assuming that tidal disruptions decay with a power law index of -5/3, the fraction of time ${\cal F}_{\rm Edd}$ a black hole accretes at Eddington due to disruptions is
\begin{align}
{\cal F}_{\rm Edd} = 3 \times 10^{-3} M_{\rm h,6}^{3/5} \Gamma_{-2},
\end{align}
resulting in an average luminosity of
\begin{align}
\langle L_{\rm TE} \rangle &= {\cal F}_{\rm Edd} L_{\rm Edd}\nonumber\\
&= 4 \times 10^{41} M_{\rm h,6}~{\rm ergs~s}^{-1}.
\end{align}
Using the optical depth from Equation (\ref{eq:agntau}), the average momentum deposition is
\begin{align}
\langle \dot{p}_{\rm TE} \rangle &= \frac{\tau L_{\rm TE}}{c}\nonumber\\
&= 4 \times 10^{32} f_{\rm Edd} M_{\rm h, 6}^{3/5}~{\rm g~cm~s}^{-2},
\end{align}
where $f_{\rm Edd}$ is the fraction of Eddington of the {\it ambient} accretion, not of the tidal disruptions, as the debris provides a negligible amount of optical depth.
 
Jets can also be produced by tidal disruptions \citep{Bloom:2011a}, although the inferred fraction of TDEs producing jets is $f_{\rm TJ} \sim 10\%$ the rate of ``thermal'' flares associated with the formation of an accretion disk \citep{Cenko:2012b}. As in the AGN jet case, such jets are collimated, but tidal disruption jets are likely more randomly oriented in angle from event to event, with some preference for the spin axis of the black hole, and are likely broader than AGN jets due to their lower relativistic $\Gamma$ \citep{Tchekhovskoy:2014a}. With these caveats, the energy and momentum deposited by TDE jets is
\begin{align}
L_{\rm TJ}  &= 1 \times 10^{42} \Gamma_{-2} f_{\rm TJ,-1} \epsilon_{\rm TJ,-1}~{\rm ergs~s}^{-1}\\
\dot{p}_{\rm TJ}  &= \frac{ L_{\rm TJ} }{c}\nonumber\\
&=3 \times 10^{31} \Gamma_{-2} f_{\rm TJ,-1} \epsilon_{\rm TJ,-1}~{\rm g~cm~s}^{-2}.
\end{align}
Lastly, tidal disruption events may also be associated with outflows of gas as they exceed their Eddington limits \citep{Metzger:2015a,Alexander:2016a,Jiang:2016b}. In such outflows, half of the output momentum is mechanical, half radiation \citep{Jiang:2014a}, resulting in the energy and momentum deposition of
\begin{align}
L_{\rm TO}  &= 7 \times 10^{42} \Gamma_{-2} \left(\frac{f_{\rm Edd}}{0.5}\right)\epsilon_{\rm TO,-1}~{\rm ergs~s}^{-1}\\
 \dot{p}_{\rm TO} &= \frac{ L_{\rm TO} }{c/3}\nonumber\\
&= 7 \times 10^{32} \Gamma_{-2} \left(\frac{f_{\rm Edd}}{0.5}\right)\epsilon_{\rm TO,-1}~{\rm g~cm~s}^{-2},
\end{align}
where $f_{\rm Edd}$ is the fraction of events that exceed Eddington at their peak, roughly 50\% of events for $M_{\rm h} \sim 10^{7} M_{\odot}$ \citep{Guillochon:2015b}.

It is clear by examining the expressions above that the UDRs beat out the other mechanisms for momentum driving by tidal disruptions primarily because the typical velocity of the effluence is $\sim 3\%~c$ as compared to $c$ and $c/3$ for jets and outflows respectively. Momentum injected by the radiation produced from the accretion of disruption debris would become competitive with the UDRs once $M_{\rm h} \gtrsim 3 \times 10^{7} M_{\odot}$, but only when the black hole is already accreting near the Eddington limit from the ambient gas surrounding the black hole. Additionally, jets and outflows likely only occur for a subset of tidal disruption events, whereas UDRs are produced with each disruption. We thus conclude that {\it UDRs are likely to be the primary form of momentum injection from tidal disruptions}.

\section{A Black Hole Disruption Engine}\label{sec:engine}
The equilibrium state of molecular clouds in a galactic nucleus are set by their thermal pressure, turbulent pressure, self-gravity, and external pressure. While the detailed dynamics of these clouds are complicated \citep{Chen:2016a}, clouds will remain in roughly virial balance and not form any stars so long as their (turbulent) Jeans mass is greater than their own mass. If a cloud of a fixed mass is supported by turbulence, as most molecular clouds are thought to be, it can begin to form stars once the turbulent energy decays to the point that the Jeans criterion is satisfied, at which time the cloud will undergo monolithic collapse and form stars. A fraction of these stars will be massive and will inject energy into the ISM via SNe.

Star formation can be indefinitely suppressed in these clouds if the turbulent energy is sustained. In order to maintain this condition, the turbulent energy must be resupplied on a timescale that is shorter than the timescale of dissipation of the turbulence, which is roughly the crossing time of the cloud $\tau_{\rm c}$. This means that an outside source of momentum must be frequent enough to deposit new momentum into a cloud before a time $\tau_{\rm c}$ has elapsed.

\begin{figure}[t]
\centering\includegraphics[width=0.9\linewidth,clip=true]{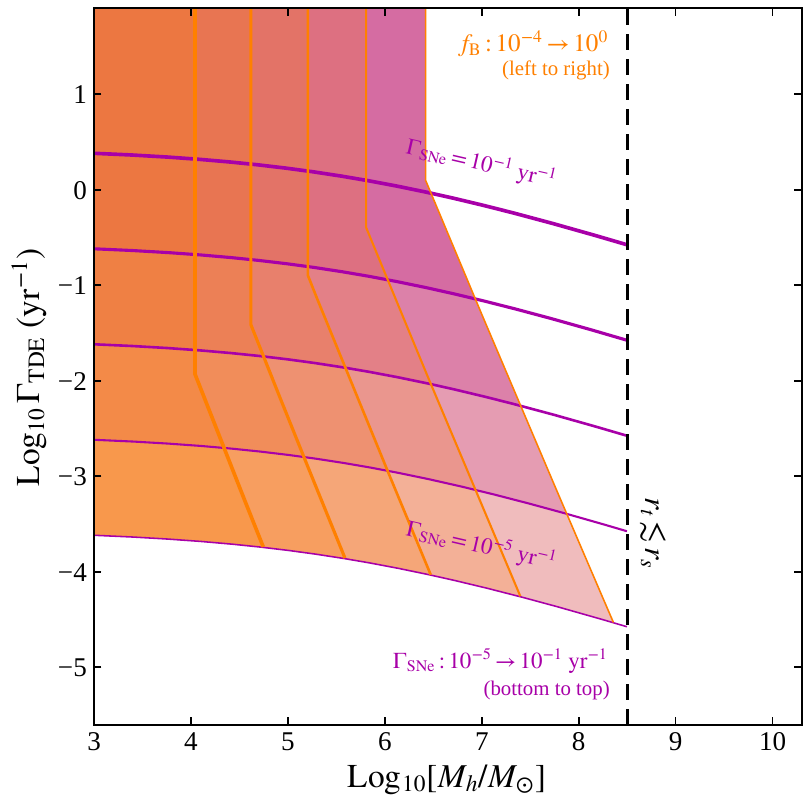}
\caption{Conditions by which TDEs dominate nuclear feedback for a variety of Bondi accretion efficiencies $f_{\rm B}$ and local SNe rates $\Gamma_{\rm SNe}$, with $f_{\rm B}$ varying from $10^{-4}$ to 1 from left to right and $\Gamma_{\rm SNe}$ varying from $10^{-5}$~yr$^{-1}$ to $10^{-1}$~yr$^{-1}$ from bottom to top (both in decade increments). Because the Eddington ratio is tied to the disruption rate through the equilibrium relation, the AGN boundaries steepen and become vertical once $f_{\rm Edd}$ saturates, which is why the orange family is ordered left to right rather than top to bottom. Within the orange (purple) regions, TDEs dominate AGN (SNe) feedback for the given values of $f_{\rm B}$ and $\Gamma_{\rm SNe}$. The vertical black dashed line indicates the limit where solar-mass stars are swallowed whole for non-spinning black holes before losing any mass via tides.}
\label{fig:rates}
\end{figure}

Comparing the most dominant subprocesses of Sections \ref{subsec:sne}, \ref{subsec:agn}, and \ref{subsec:tde} shows that if the tidal disruption rate is large, the momentum that TDEs provide can dominate all other forms of feedback, with the UDRs resulting from the unbound ejecta being the most important mechanism among the disruption feedback processes. By fixing the supernova rate and AGN accretion rates to fixed values, Figure \ref{fig:rates} shows the combinations of parameters for which tidal disruptions become the primary source of feedback for the MCs surrounding a SMBH of various masses. The figure shows that for a disruption rate $\Gamma_{\rm TDE} \sim 10^{-2}$, tidal disruptions beat out both AGN and SNe feedback processes provided the AGN is only modestly active and the SNe rate does not exceed a value comparable to the rate within the MW's central kpc, $\Gamma_{\rm SNe} \sim 10^{-4}~{\rm yr}^{-1}$ (see Equation (\ref{eq:gammasne})).

\subsection{The collapse of externally-driven MCs and the underlying nucleus}

Assuming now that TDEs are primarily responsible for driving turbulence in the MCs surrounding the nucleus, the clouds will collapse gravitationally until they are in a new equilibrium where their enhanced level of turbulent support balances with their self-gravity. Because of the negative specific heat of self-gravitating systems \citep{Lynden-Bell:1968a}, the kinetic ``heating'' causes the density of the clouds to increase significantly (Figure \ref{fig:diagram}). Now denser, the MCs are free to dive deeper into the nuclear stellar cluster via dynamical friction without risk of tidal disruption by the black hole's gravity.

As the MCs burrow into the nucleus' outer layers, they proceed to scatter stars near them. This process removes the stars of lower binding energy, with the black hole only being able to retain the more-tightly bound stars closer to it. In an equilibrium state, the nuclear cluster surrounding the SMBH should have a mass of stars comparable to the mass of the SMBH. Restoring this equilibrium requires the black hole to capture stars of higher orbital energy that will not be removed by the MCs scouring its outer layers. Again, the negative specific heat of the system comes into play: In response to the frictive heating provided by the infalling MCs, the entire nucleus {\it also shrinks} until a new equilibrium is established. As we show in Section~\ref{sec:equil}, the clouds do not come to rest at the sphere of influence itself, but settle at a radius roughly two orders of magnitude larger, where they can survive the black hole's tidal field while still dominating the relaxation of the nucleus.

This effect is visible in N-body simulations of \citet{Antonini:2012a} (see their Figure 13) in which a massive body plunges into a SMBH cluster; the cluster shrinks in response to the presence of the sinking body. In the scenario proposed here, dozens of massive MCs drag into the nuclear cluster simultaneously, the cumulative effect of which compresses the cluster. We note in advance that the compression is not one that drives the stellar density down to the cloud density: the solution of Section~\ref{sec:equil} places the clouds at $r_{\rm b} \sim 10^{2} a_{\rm h}$, far outside the cusp, and leaves the cusp some five orders of magnitude denser than the clouds that perturb it. What the clouds supply is not a density but a relaxation rate, and we make that statement quantitative in Section~\ref{sec:compcond}.

\subsection{An increased rate of disruptions and collisions}\label{subsec:increase}

With a higher stellar density and higher velocity dispersion, the two-body relaxation rate of the nuclear cluster shrinks dramatically, driving more stars into the loss cone and increasing the tidal disruption rate. These new disruptions further increase the rate of turbulent driving into the surrounding MCs, increasing their density, and thus further enhancing the rate. Without another process to halt the runaway, the stellar disruption rate could in principle become infinitely large.

Because the interactions between stars increases significantly, the rate of star-star collisions also increases as the nuclear cluster collapses. As we shall see in Section \ref{sec:collcond}, the rate of star-star collisions depends on the stellar density $n_{\ast}$, but the rate of tidal disruptions depends on the properties of the surrounding dense MCs. As a result, the rate of stellar collisions increases more rapidly than the rate of tidal disruptions, and at some point the probability that a star strikes another star becomes comparable to the rate it will be disrupted by the central black hole. 

As the stellar densities are largest closest to the black hole, most of the collisions occur just outside the tidal radius, the distance within which no stars exist. A star on its way to disruption would likely be scattered there in a single orbit about a normal nuclear cluster, but the compression of the core above the usual values means that even main-sequence stars are likely to be in the diffusion regime of disruption \citep{Wang:2004a,MacLeod:2012a}. In the diffusion regime, the stars spend many orbital periods in an orbit similar to their eventual disruptive orbit; each of these orbits presents the possibility of a stellar collision.

A collision between two stars effectively randomizes the orbits of the two stars (if they remain separate) or their merger product, removing them from the loss cone, and saving the star from being disrupted by the SMBH. This effectively caps the tidal disruption rate to a value no greater than the probability of stellar collision.

\section{Equilibrium state of TDE engines}\label{sec:equil}

A steady-state disruption engine requires the process to have both positive and negative feedback loops; if the disruption rate increases above its equilibrium value, some other process should act to reduce it back to this value, and vice versa if the disruption rate drops below this value. As discussed in the previous sections, disruptions enhance the nuclear density leading to more disruptions (the positive side of the loop), whereas collisions will prevent disruptions from exceeding a critical value (the negative side of the loop).

\subsection{Structure of the nucleus and the surrounding cloud belt}\label{subsec:structure}

We describe the equilibrium with four quantities: the velocity dispersion $\sigma$ that sets the size of the sphere of influence, the radius $r_{\rm b}$ at which the molecular clouds orbit, their internal Mach number ${\cal M}$, and their radius $R_{\rm MC}$. Three physical conditions, described in the following section, close the system, leaving $M_{\rm h}$ (and the Bondi efficiency $f_{\rm B}$, which enters only the derived accretion quantities) as the sole free parameter. A fourth condition, developed in Section~\ref{sec:compcond}, does not enter the closure but must hold as an inequality for the solution to be self-consistent; it asks whether the clouds are in fact capable of compressing the cusp to the state the other three imply, and it is what sets the range of $M_{\rm h}$ over which an engine can exist.

The stellar cluster is taken to be a relaxed Bahcall-Wolf cusp, $n_{\ast}(r) \propto r^{-\gamma}$ with $\gamma = 7/4$, containing a mass in stars equal to twice the black hole mass within its sphere of influence,
\begin{equation}
a_{\rm h} = \frac{2 G M_{\rm h}}{\sigma^{2}} = 1.5 \mhsix \sigma_{75}^{-2}~{\rm pc},\label{eq:ah}
\end{equation}
where $\sigma_{75}$ is $\sigma$ in units of $75~{\rm km~s}^{-1}$. With an average stellar mass $m_{\ast} = 0.5~\msun$ and radius $R_{\ast} = 0.46~R_{\odot}$ \citep{Tout:1996a}, the number of stars is $N_{\ast} = 2 M_{\rm h}/m_{\ast}$ and the density normalization of the cusp is $n_{0} = (3-\gamma)N_{\ast}/4\pi a_{\rm h}^{3}$. Within the cusp the one-dimensional dispersion is $\sigma_{\rm h} = \sigma/\sqrt{2}$.

The molecular clouds cannot themselves reside at $a_{\rm h}$. This is a departure from the picture sketched in Section~\ref{sec:engine}, and it is forced on us by the solution: at the compressed densities the cluster reaches, a cloud placed at $a_{\rm h}$ would need a density $\gtrsim 10^{10}~{\rm cm}^{-3}$ to survive the black hole's tidal field, at which point no UDR could drive it (its radiative length would fall below the cloud's own size), and the first of our conditions below has no solution. The clouds instead settle into a belt at a radius $r_{\rm b} \gg a_{\rm h}$, where the tidal limit is far less severe. Structures of exactly this kind are now routinely resolved: ALMA imaging of NGC~1068 reveals a rotating molecular disk of $\sim 10^{5}~\msun$ and $7$--$10$~pc diameter surrounding the black hole \citep{GarciaBurillo:2016a,Imanishi:2018a,Impellizzeri:2019a}, and surveys of nearby Seyferts find such compact molecular disks to be the norm rather than the exception, with median diameters of tens of parsecs and masses of a few $\times 10^{5}~\msun$ \citep{Combes:2019a,GarciaBurillo:2021a}. Our own Galactic center hosts the same structure at smaller scale in its circumnuclear disk. The radius and mass we derive below are set by the equilibrium conditions rather than fitted, and it is worth stating in advance that they land close to these observed values. There they are marginally tidally bound,
\begin{equation}
\rho_{\rm MC} = \frac{9 M_{\rm h}}{4 \pi r_{\rm b}^{3}},\label{eq:rhomc}
\end{equation}
and supported against their own self-gravity by supersonic turbulence,
\begin{equation}
{\cal M}^{2} c_{\rm s}^{2} = \frac{4 \pi}{5} G \rho_{\rm MC} R_{\rm MC}^{2},\label{eq:virial}
\end{equation}
with $c_{\rm s} = 0.19~{\rm km~s}^{-1}$ the isothermal sound speed of $10$~K molecular gas at $\mu = 2.33$. The turbulent velocity within each cloud is $\sigma_{\rm cl} = {\cal M} c_{\rm s}$ and the crossing time on which that turbulence decays is $\tau_{\rm c} = 2 R_{\rm MC}/\sigma_{\rm cl}$. The number density of gas in a cloud is $n_{\rm MC} = \rho_{\rm MC}/1.4 m_{\rm p}$, the mass of each cloud is $M_{\rm MC} = 4 \pi \rho_{\rm MC} R_{\rm MC}^{3}/3$, and $N_{\rm MC} = 4 r_{\rm b}^{2}/R_{\rm MC}^{2}$ clouds are required to cover the belt.

That the clouds sit well outside the cusp is what allows them to keep the loss cone full while remaining dynamically benign for the cusp itself, as we discuss in Section~\ref{sec:collcond}.

\subsection{Conditions for equilibrium}\label{subsec:conditions}

\subsubsection{Finite sizes of UDRs}

While more momentum can be driven into the MCs by the UDRs than any other process, it is not necessarily the case that this momentum will be shared equally among all MCs surrounding the SMBH. Each UDR will only be in contact with a size scale comparable to its Sedov-Taylor length $R_{\rm ST}$, which is small compared to the belt radius. As all MCs must be struck at least once per crossing time to sustain their turbulent energy, multiple tidal disruptions are required to cover the full $4\pi$ steradians of solid angle surrounding the black hole. This sets our first condition,
\begin{equation}
\Gamma_{\rm TDE} f_{\rm UDR} \tau_{\rm c} = 1,\label{eq:cond1}
\end{equation}
where $f_{\rm UDR} = R_{\rm UDR}^{2}/4 r_{\rm b}^{2}$ is the fractional solid angle occupied by a single UDR. This fraction depends on the Sedov-Taylor length \citep{Sedov:1946a,Taylor:1950a} within the clouds the UDRs are striking,
\begin{equation}
R_{\rm ST} = 1.15 \left(\frac{E}{\rho_{\rm MC}}\right)^{1/5} \tau_{\rm rad}^{2/5},\label{eq:rst}
\end{equation}
where $\tau_{\rm rad}$ is the timescale for the UDR to become radiative, itself set by the density of the clouds being impacted,
\begin{equation}
\tau_{\rm rad} = 130 E_{50}^{4/17}n_{\rm MC,4}^{-9/17}~{\rm yr},\label{eq:trad}
\end{equation}
following the numerical SNR calculations of \citet{Blondin:1998a}. Assuming a solar composition and combining Equations (\ref{eq:rst}) and (\ref{eq:trad}), we find
\begin{equation}
R_{\rm UDR} = 0.22 E_{50}^{5/17} n_{\rm MC,4}^{-7/17}~{\rm pc},\label{eq:rudr}
\end{equation}
evaluated at the cloud density $n_{\rm MC}$. At this distance the remnant begins to radiate away much of its energy rather than depositing it into the surrounding MCs, and it is at this time that the majority of the momentum is deposited as the remnant sweeps up a mass several times its own. Note that these expressions presume the UDR solely interacts with the cold MC gas phase and not the surrounding hot phase; because of its much lower density, a UDR will propagate to a much larger size before becoming radiative in the hot phase. However, the unbound debris streams that produce UDRs will almost certainly strike a dense MC before reaching their termination distance in the surrounding hot ISM \citep{Guillochon:2016b}, with the MCs having a solid angle coverage fraction of order unity even in our own MW \citep{Chen:2016b}, which is a tepid accretor.

\subsubsection{Heating and cooling of MCs}

Both momentum and energy are injected into the MCs as UDRs slam into them. As turbulence decays, the kinetic energy of the turbulence is converted into heat, which the cloud must radiate away at the same rate at which energy is being injected. This defines a second condition,
\begin{equation}
\Gamma_{\rm TDE} E_{\rm UDR} = L_{\rm MC},\label{eq:cond2}
\end{equation}
where $E_{\rm UDR} = 1.8 \times 10^{50} \mhsix^{1/3}$~ergs is the energy of the unbound debris \citep{Guillochon:2016b} and $L_{\rm MC}$ is the total cooling rate of all MCs in the belt. Because the rate of cooling is a steep function of temperature below H-recombination, the MCs rapidly cool to a temperature where their cooling rates are equal to the cosmic ray heating rate, $\sim 10~{\rm K}$ \citep{Krumholz:2015a}, with CO cooling being the predominant coolant. Writing the volumetric cooling rate as $\Lambda_{0} n_{\rm MC}^{2}$ with $\Lambda_{0} = 1.3 \times 10^{-27}~{\rm erg~cm}^{3}~{\rm s}^{-1}$ \citep{Goldsmith:1978a,Neufeld:1995a}, and summing over the $N_{\rm MC}$ clouds of the belt,
\begin{equation}
L_{\rm MC} = \frac{16 \pi}{3} \Lambda_{0} n_{\rm MC}^{2} r_{\rm b}^{2} R_{\rm MC}.\label{eq:lmc}
\end{equation}

\subsubsection{Stellar collisions preventing a disruption runaway}\label{sec:collcond}

As described in Section \ref{subsec:increase}, the rate of star-star collisions in an isotropic cluster increases as the density of the cluster increases during the collapse of the nuclear cluster. For a star to be successfully disrupted, it must run the gauntlet of the high-density central region of the cluster where it is at risk of colliding with another star. This yields our last condition,
\begin{equation}
\Gamma_{\rm coll} = \Gamma_{\rm TDE}.\label{eq:cond3}
\end{equation}
The two rates on either side of this expression must be evaluated with some care, and both differ from the estimates used in earlier treatments of this problem.

Because the stellar density rises steeply toward the center, we obtain $\Gamma_{\rm coll}$ by integrating the collision rate over the cusp rather than evaluating it at a single radius. Crucially, at the densities of interest the orbital speeds within most of the cusp remain below the escape speed from a star's own surface, $v_{\rm esc,\ast} = 644~{\rm km~s}^{-1}$, so collisions are gravitationally focused and the geometric cross-section badly underestimates the rate. Retaining the focusing term, integration of $n_{\ast}^{2} \Sigma_{\rm coll} v$ over the $\gamma = 7/4$ cusp gives
\begin{equation}
\Gamma_{\rm coll} = 8 \pi^{2} \Lambda_{\rm c} R_{\ast}^{2} v_{\rm esc,\ast}^{2} n_{0}^{2} a_{\rm h}^{7/2} \left(G M_{\rm h}\right)^{-1/2},\label{eq:gcoll}
\end{equation}
where $\Lambda_{\rm c} \simeq 3$ collects the dimensionless integral over the cusp profile. Neglecting focusing, as is sometimes done, moves the collisional cap to velocity dispersions that are not physically attainable.

For the disruption rate we must be careful about which loss-cone regime the cusp occupies. The compressed clusters we derive below turn out to sit close to the boundary between the two: the per-orbit angular momentum kick is comparable to the width of the loss cone, so neither the full-loss-cone (pinhole) limit nor the empty (diffusive) limit applies cleanly. We therefore use the bridging flux of \citet{Cohn:1978a}, in the form given by \citet{Merritt:2013a} and \citet{Vasiliev:2013a},
\begin{equation}
{\cal F}(\epsilon) = \frac{4 \pi^{2} J_{\rm c}^{2}(\epsilon)\, f(\epsilon)\, \bar{\mu}(\epsilon) P(\epsilon)}{\ln\left[1/{\cal R}_{0}(\epsilon)\right]},
\label{eq:gtde}
\end{equation}
where ${\cal R}_{\rm lc} = J_{\rm lc}^{2}/J_{\rm c}^{2} = 4\epsilon r_{\rm t}/GM_{\rm h}$, $q(\epsilon) = \bar{\mu}P/{\cal R}_{\rm lc}$, and ${\cal R}_{0} = {\cal R}_{\rm lc}\exp[-\alpha(q)]$ with $\alpha(q) = (q^{4}+q^{2})^{1/4}$. The orbit-averaged diffusion coefficient $\bar{\mu}(\epsilon)$, which arises from $\langle(\Delta v_{\perp})^{2}\rangle$, is evaluated in closed form for a power-law cusp in a Keplerian potential using Appendix A of \citet{Stone:2016a}. The disruption rate is $\Gamma_{\rm TDE} = \int {\cal F}(\epsilon)\,{\rm d}\epsilon$ over the cusp. Equation~(\ref{eq:gtde}) reduces to $4\pi^{2}J_{\rm c}^{2}f{\cal R}_{\rm lc}$ for $q \gg 1$ and to the diffusive flux for $q \ll 1$; we have verified both limits numerically to four significant figures.

That Equation~(\ref{eq:gtde}) is not the pinhole rate has a direct consequence. In the pinhole limit $\Gamma_{\rm TDE}$ and $\Gamma_{\rm coll}$ carry the {\it same} power of $\sigma$ once the closure $n_{0} \propto M_{\rm h}/a_{\rm h}^{3}$, $a_{\rm h} = 2GM_{\rm h}/\sigma^{2}$ is imposed, $\sigma$ cancels identically from Equation~(\ref{eq:cond3}), and no equilibrium exists at all. The degeneracy is broken only by the $q$-dependence of Equation~(\ref{eq:gtde}) --- that is, by how far into the diffusive regime the cusp has been driven. The equilibrium is in this sense a statement about the degree of loss-cone depletion, and it necessarily lies near $q \sim 1$.

Applied to nuclei with observed properties rather than to the engine, Equation~(\ref{eq:gtde}) returns $6 \times 10^{-5}$~yr$^{-1}$ for a Milky Way-like nucleus ($M_{\rm h} = 4.3\times10^{6}~\msun$, $M(<1\,{\rm pc}) = 10^{6}~\msun$, $\gamma = 7/4$), within the range of standard loss-cone calculations, with $q \simeq 1.5$ at $1$~pc; and $\Gamma_{\rm TDE} \propto M_{\rm h}^{-0.15}$ along the $M$--$\sigma$ relation, to be compared with the $M_{\rm h}^{-0.4}$ of \citet{Stone:2016a}, the difference reflecting our use of a single fixed $\gamma$ and no stellar mass function.

Equation~(\ref{eq:gcoll}) and the integral of Equation~(\ref{eq:gtde}) both depend only on $\sigma$ once $M_{\rm h}$ is fixed, so Equation~(\ref{eq:cond3}) alone determines the compression of the cluster; the remaining two conditions then fix the properties of the belt.

\subsubsection{Can the clouds deliver the compression?}\label{sec:compcond}

Conditions (\ref{eq:cond1})--(\ref{eq:cond3}) close the system, but they do not by themselves verify the positive branch of the loop. Equation~(\ref{eq:cond3}) fixes $\sigma$ from the cusp alone, and Equations~(\ref{eq:cond1}) and (\ref{eq:cond2}) then determine what the clouds must look like; nothing in that sequence asks whether clouds of those properties are capable of compressing the cusp to that $\sigma$ in the first place. Since that compression is the mechanism the model rests on, we impose it as a separate requirement.

The clouds do not reach the cusp --- the tidal condition of Equation~(\ref{eq:rhomc}) strands them at $r_{\rm b} \sim 10^{2} a_{\rm h}$ --- so the appropriate statement is not that they raise the stellar density to their own, but that they dominate the relaxation of the cusp, and hence the rate at which it evolves. Writing $t_{\rm relax,MC} = 0.34\sigma^{3}/(G^{2} M_{\rm MC} \rho_{\rm MC} \ln\Lambda)$ for the relaxation time driven by the clouds as massive perturbers, the marginal case is
\begin{equation}
t_{\rm relax,MC} = t_{\rm relax,\ast}.\label{eq:cond4}
\end{equation}
Beyond this point the cusp relaxes faster on its own than the clouds can drive it, and further compression cannot be attributed to them; we discuss the limitations of Equation~(\ref{eq:cond4}) in Section~\ref{subsec:caveats}. Replacing Equation~(\ref{eq:cond3}) with Equation~(\ref{eq:cond4}) and re-solving the same monomial system gives the largest dispersion the clouds can sustain,
\begin{equation}
\sigma_{\rm comp} = 2.8 \times 10^{2} \mhsix^{0.40}~{\rm km~s}^{-1},\label{eq:sigcomp}
\end{equation}
essentially independent of geometry (we obtain $275$ and $272~{\rm km~s}^{-1}$ for the isotropic and disk-fed cases respectively, the small difference entering only through the cloud count in Equation~(\ref{eq:lmc})).

The engine settles at whichever of the two dispersions is smaller. Where $\sigma_{\rm comp} > \sigma_{\rm coll}$ the clouds are more than able to supply the compression, the collision cap is the binding constraint, and the solution of the following section stands unmodified. Where $\sigma_{\rm comp} < \sigma_{\rm coll}$ the clouds are the bottleneck, collisions never become important, and both $\sigma$ and $\Gamma$ fall below the values quoted below. Because $\sigma_{\rm comp} \propto \mhsix^{0.40}$ rises more steeply with mass than $\sigma_{\rm coll} \propto \mhsix^{1/6}$, the criterion takes the form of a lower bound on the black hole mass,
\begin{equation}
\mhsix \gtrsim 8.9\, f_{\ast}^{1.64}.\label{eq:mcrit}
\end{equation}
Two features of the underlying rates make Equation~(\ref{eq:mcrit}) this simple. The dispersion the collision cap permits, $\sigma_{\rm coll} = 6.5 \times 10^{2} f_{\ast}^{0.66}~{\rm km~s}^{-1}$, is independent of black hole mass to within its fitted exponent ($\mhsix^{b}$ with $b = 0.001$--$0.044$ across $0.14 \le f_{\ast} \le 1$), so we drop the mass term; the entire mass dependence of the criterion therefore enters through $\sigma_{\rm comp} \propto \mhsix^{0.40}$. Equation~(\ref{eq:mcrit}) reproduces the numerically solved floor to better than $7\%$ over the range of interest, and gives $M_{\rm h} > 6 \times 10^{6}~\msun$ for an isotropic cusp against only $M_{\rm h} > 6 \times 10^{5}~\msun$ for the disk-fed cusp we prefer. This is a further, and independent, reason to favor the disk-fed geometry of Section~\ref{subsec:disk}: the isotropic solution fails its own compression requirement over most of the mass range for which it is quoted, whereas the disk-fed solution satisfies it comfortably.

\subsection{The Engine State: the Isotropic Limit}\label{subsec:engine-state}

{\it The solution presented in this section is a limiting case, and is not the configuration we advocate.} We solve first for an isotropic cusp, $f_{\ast} = 1$, because it is the simplest case and the one in which the structure of the equilibrium is clearest. It is not, however, a configuration the model can actually occupy: Section~\ref{subsec:window} shows that $f_{\ast} = 1$ satisfies neither the compression criterion of Section~\ref{sec:compcond} nor the requirement that tidal debris outweigh AGN feedback, and the cusp it demands is far denser than any observed nuclear star cluster. The reader interested only in the physical solution should turn to Section~\ref{subsec:disk}, and to the admissible range established in Section~\ref{subsec:window}. We give the isotropic case in full because every later result is obtained by the same construction, and because the contrast between it and the disk-fed case is instructive.

The three conditions (Equations (\ref{eq:cond1}), (\ref{eq:cond2}), and (\ref{eq:cond3})) are each monomial in the four unknowns, and so the system can be solved in closed form by taking logarithms; the solution is a set of exact power laws in $M_{\rm h}$, which we quote below to two significant figures. For the nuclear cluster we find
\begin{align}
\sigma &= 6.4 \times 10^{2}~{\rm km~s}^{-1}\\
a_{\rm h} &= 2.1 \times 10^{-2} \mhsix~{\rm pc}\\
n_{0} &= 4.4 \times 10^{10} \mhsix^{-2}~{\rm pc}^{-3}\\
P_{\rm h} &= 2.8 \times 10^{2} \mhsix~{\rm yr}\\
\tau_{\rm relax,\ast} &= 4.3 \times 10^{7} \mhsix^{2}~{\rm yr},
\end{align}
where $P_{\rm h}$ is the orbital period at $a_{\rm h}$ and $\tau_{\rm relax,\ast}$ the two-body relaxation time of the cusp. The properties of the belt and of the clouds within it are
\begin{align}
r_{\rm b} &= 1.9 \mhsix^{1.08}~{\rm pc}\\
{\cal M} &= 19 \mhsix^{0.53}\\
R_{\rm MC} &= 0.10 \mhsix^{1.65}~{\rm pc}\\
n_{\rm MC} &= 3.2 \times 10^{6} \mhsix^{-2.23}~{\rm cm}^{-3}\\
\sigma_{\rm cl} &= 3.6 \mhsix^{0.53}~{\rm km~s}^{-1}\\
\tau_{\rm c} &= 5.6 \times 10^{4} \mhsix^{1.12}~{\rm yr}\\
M_{\rm MC} &= 5.2 \times 10^{2} \mhsix^{2.71}~\msun\\
N_{\rm MC} &= 1.3 \times 10^{3} \mhsix^{-1.14}\\
M_{\rm belt} &= 6.7 \times 10^{5} \mhsix^{1.57}~\msun,
\end{align}
corresponding to a gas surface density through the belt of $6.2 \times 10^{4} \mhsix^{-0.58}~\msun~{\rm pc}^{-2}$. The equilibrium disruption rate is
\begin{equation}
\Gamma_{\rm eq} = \Gamma_{\rm coll} = 4.2 \times 10^{-1} \mhsix^{-1.00}~{\rm yr}^{-1},\label{eq:eqrate}
\end{equation}
a factor of several thousand above the canonical $10^{-4}$~yr$^{-1}$ at $10^{6}~\msun$, and falling almost inversely with black hole mass. The size, radiative time, angular coverage, and collective luminosity of the UDRs are
\begin{align}
R_{\rm UDR} &= 2.4 \times 10^{-2} \mhsix~{\rm pc}\\
\tau_{\rm rad} &= 7.0 \mhsix^{1.26}~{\rm yr}\\
f_{\rm UDR} &= 4.2 \times 10^{-5} \mhsix^{-0.12}\\
L_{\rm UDR} = L_{\rm MC} &= 2.4 \times 10^{42} \mhsix^{-0.66}~{\rm erg~s}^{-1}.
\end{align}
The density of the gas surrounding the black hole, the resulting accretion luminosity, and the typical extinction through the ambient medium are
\begin{align}
n_{\rm bg} &= 1.8 \times 10^{4} \mhsix^{-1.09}~{\rm cm}^{-3}\\
L_{\rm AGN}/L_{\rm Edd} &= 3.0 \times 10^{-2} f_{\rm B,-3} \mhsix^{0.02}\\
A_{\rm V} &\simeq 47 \mhsix^{-0.02},
\end{align}
and finally the average feeding rate of the black hole from TDEs relative to accretion of the surrounding gas is
\begin{equation}
\langle \dot{M}_{\rm TDE} \rangle / \dot{M}_{\rm amb} = 79 f_{\rm B,-3}^{-1}\mhsix^{-2.02}.\label{eq:eqaccratio}
\end{equation}

Several features of this solution deserve comment. The equilibrium dispersion is essentially independent of black hole mass, so that $\Gamma_{\rm eq} \propto \mhsix^{-1}$ and the rate falls by four decades across the four decades of $M_{\rm h}$ over which the engine can operate: the engine is strongly a low-mass phenomenon. The clouds are also extreme in this limit, with $n_{\rm MC} \sim 3 \times 10^{6}~{\rm cm}^{-3}$ and radii of $\sim 0.1$~pc, denser and more compact than anything observed in a galactic nucleus. Both features are consequences of the isotropic assumption rather than of the engine itself, and both are substantially relaxed in the disk-fed solution of Section~\ref{subsec:disk}, where the clouds become close analogues of those in the Milky Way's own Central Molecular Zone, of which G0.253+0.016 (``the Brick'') is the best-studied example \citep{Longmore:2013a,Kruijssen:2014a,Barnes:2017a}. The engine state, in other words, does not require gas in an unobserved configuration provided the cusp is flattened; it requires CMZ-like gas around a nucleus in which the disruption rate has already been raised.

The extinction is the most immediately testable consequence: $A_{\rm V} \simeq 50$ means that an engine nucleus is optically obscured, and its disruptions would be missed entirely by optical surveys while remaining detectable in the infrared.

We also note the assumptions that limit the range of validity of Equations (\ref{eq:eqrate})--(\ref{eq:eqaccratio}). As noted above, the isotropic case is a limiting one: the compression criterion of Section~\ref{sec:compcond} imposes a mass floor of $8 \times 10^{6}~\msun$ on it, and its central density, $\rho_{0} \simeq 2 \times 10^{10}~\msun~{\rm pc}^{-3}$, exceeds any observed nuclear star cluster by several orders of magnitude. The gravitational focusing that sets Equation~(\ref{eq:gcoll}) requires $\sigma_{\rm h} < v_{\rm esc,\ast}$, which fails for $M_{\rm h} \gtrsim 10^{8}~\msun$; above $10^{8.5}~\msun$ main-sequence stars are swallowed whole in any case. The equilibrium sits near $q_{\ast} \sim 1$, at the boundary between the full and empty loss cone regimes, which is why Equation~(\ref{eq:gtde}) rather than a limiting form is required; the numerical value of $\Gamma_{\rm eq}$ is correspondingly sensitive to the collision normalization $\Lambda_{\rm c}$, varying by two orders of magnitude for $1 \le \Lambda_{\rm c} \le 10$, while the mass exponent is stable to within $\pm 0.1$. Rates should be read with that uncertainty attached. We note finally that the belt clouds cannot themselves torque cusp stars efficiently; the loss cone is more plausibly kept populated by stars formed in the belt clouds on low-angular-momentum orbits, which is the resupply channel implied by the association of TDEs with recent star formation, though we have not estimated its rate and it sits in tension with the suppression of star formation that Condition (\ref{eq:cond1}) is designed to achieve.

\subsection{Disk geometry and the direction of the debris}\label{subsec:disk}

The solution above treats the clouds as an isotropic belt covering $4\pi$ steradians. This is almost certainly wrong in detail. Gas that settles in a galactic nucleus carries angular momentum, and the structures actually resolved around nearby black holes are disks: the NGC~1068 torus is a rotating molecular disk a few parsecs across \citep{GarciaBurillo:2016a,Imanishi:2018a,Impellizzeri:2019a}, and compact molecular disks of comparable size and mass are found in most nearby Seyferts surveyed \citep{Combes:2019a,GarciaBurillo:2021a}. We therefore let the clouds occupy a disk of aspect ratio
\begin{equation}
f_{\Omega} \equiv H/r_{\rm b},
\end{equation}
which is the fraction of $4\pi$ the disk subtends as seen from the black hole. Observed circumnuclear disks are geometrically thick, with $f_{\Omega} \sim 0.1$--$0.3$; we adopt $f_{\Omega} = 0.2$ as a fiducial value and quote the scalings explicitly.

\subsubsection{Why the isotropic solution survives an isotropic cusp}

It is worth noting first that if the stellar cusp remains isotropic, flattening the gas into a disk changes nothing at all; two factors of $f_{\Omega}$ enter and cancel: the number of clouds needed to cover the structure drops to $N_{\rm MC} = 4 f_{\Omega} r_{\rm b}^{2}/R_{\rm MC}^{2}$, reducing the total cooling rate by $f_{\Omega}$; but only a fraction $f_{\Omega}$ of isotropically directed UDRs strike the disk at all, reducing the heating rate by the same factor. Condition (\ref{eq:cond2}) is therefore unchanged. The same cancellation occurs in condition (\ref{eq:cond1}): there is less sky to cover, but proportionally fewer remnants doing the covering, and the covering time $4 f_{\Omega} r_{\rm b}^{2}/(\Gamma f_{\Omega} R_{\rm UDR}^{2})$ is independent of $f_{\Omega}$.

Every quantity in Section~\ref{subsec:engine-state} thus carries over unmodified, with the single exception of the total gas mass, which becomes
\begin{equation}
M_{\rm disk} = f_{\Omega} M_{\rm belt} = 4.8 \times 10^{5} \left(\frac{f_{\Omega}}{0.2}\right)\mhsix^{1.14}~\msun.
\end{equation}
This is a welcome change: $2.4 \times 10^{6}~\msun$ of dense gas within a few parsecs is uncomfortably large, whereas $\sim 5 \times 10^{5}~\msun$ sits squarely among the molecular disk masses measured by \citet{GarciaBurillo:2021a}, whose sample has a median of $\sim 6 \times 10^{5}~\msun$.

\subsubsection{The cusp and debris are both anisotropic}

The isotropic assumption is however inconsistent with the resupply channel we invoked in Section~\ref{sec:collcond}. If the loss cone is kept populated by stars forming in the belt clouds, those stars are born in the disk plane and arrive at the black hole from it. The Milky Way shows exactly this: the young stars within $0.5$~pc of Sgr~A$^{\ast}$ occupy a thin, coherently rotating disk that is most naturally understood as the fragmented remnant of a nuclear gas disk \citep{Levin:2003a,Paumard:2006a,Yelda:2014a}. On larger scales, M31's double nucleus is a stellar disk of apsidally aligned orbits \citep{Tremaine:1995a}, and such eccentric nuclear disks are now thought to be common in post-merger nuclei \citep{Madigan:2018a,Akiba:2021a}.

The direction of the ejecta follows immediately. A star disrupted from the sphere of influence is on a very nearly parabolic orbit, and the incoming and outgoing asymptotes of a parabola are parallel: the star leaves along the direction from which it came. The unbound debris is therefore launched back toward the star's point of origin, and if stars are fed from the disk, {\it the UDRs are beamed into the disk} rather than escaping isotropically into the hot phase. Writing $b$ for the fraction of UDRs that intercept the clouds, the isotropic case is $b = f_{\Omega}$ and the disk-fed case is $b \simeq 1$. Conditions (\ref{eq:cond1}) and (\ref{eq:cond2}) depend on geometry only through the ratio $\beta \equiv b/f_{\Omega}$,
\begin{align}
\Gamma_{\rm TDE} f_{\rm UDR} \tau_{\rm c}\, \beta &= 1\\
\Gamma_{\rm TDE} E_{\rm UDR}\, b &= \frac{16 \pi}{3} f_{\Omega} \Lambda_{0} n_{\rm MC}^{2} r_{\rm b}^{2} R_{\rm MC},
\end{align}
with $\beta = 1$ recovering the isotropic solution and $\beta = f_{\Omega}^{-1}$ describing beamed driving.

The collision cap is affected as well. Collisions are a local-density process, so confining the same $N_{\ast}$ stars to a fraction $f_{\ast}$ of $4\pi$ raises the density by $1/f_{\ast}$ and lowers the occupied volume by $f_{\ast}$, giving $\Gamma_{\rm coll} \rightarrow \Gamma_{\rm coll}/f_{\ast}$. The full-loss-cone disruption rate does not scale the same way: in the pinhole regime it is set by the number of stars and their orbital periods, not by the local density, and so is left unenhanced. Condition (\ref{eq:cond3}) becomes $\Gamma_{\rm coll}/f_{\ast} = \Gamma_{\rm TDE}$, and because collisions are now more efficient at fixed $\sigma$, the cap begins affecting the rates at a {\it lower} compression than in the isotropic case. Solving as before gives the closed-form dependence
\begin{align}
\sigma &= 4.0 \times 10^{2}\, f_{\ast}^{1/2}\, \mhsix^{1/6}~{\rm km~s}^{-1}\\
\Gamma_{\rm eq} &\simeq 1.4 \times 10^{-2} \left(\frac{f_{\ast}}{0.22}\right)^{2.4} \mhsix^{-0.84}~{\rm yr}^{-1}.\label{eq:gammaeqf}
\end{align}

\subsubsection{The disk-fed engine}

Taking the self-consistent case in which the stars share the geometry of the gas that formed them, $f_{\ast} = f_{\Omega} = 0.2$ and $b = 1$, the equilibrium becomes
\begin{align}
\sigma &= 2.3 \times 10^{2} \mhsix^{0.03}~{\rm km~s}^{-1}\\
\Gamma_{\rm eq} &= 1.1 \times 10^{-2} \mhsix^{-0.83}~{\rm yr}^{-1}\\
a_{\rm h} &= 0.17 \mhsix^{0.93}~{\rm pc}\\
n_{0} &= 8.3 \times 10^{7} \mhsix^{-1.80}~{\rm pc}^{-3}\\
r_{\rm b} &= 5.1 \mhsix^{0.99}~{\rm pc}\\
R_{\rm MC} &= 0.82 \mhsix^{1.48}~{\rm pc}\\
n_{\rm MC} &= 1.5 \times 10^{5} \mhsix^{-1.98}~{\rm cm}^{-3}\\
\sigma_{\rm cl} &= 6.2 \mhsix^{0.49}~{\rm km~s}^{-1}\\
M_{\rm MC} &= 1.2 \times 10^{4} \mhsix^{2.45}~\msun\\
N_{\rm MC} &= 32 \mhsix^{-0.97}\\
M_{\rm disk} &= 3.8 \times 10^{5} \mhsix^{1.48}~\msun\\
A_{\rm V} &\simeq 50,
\end{align}
with $L_{\rm AGN}/L_{\rm Edd} = 5.3 \times 10^{-2} f_{\rm B,-3} \mhsix^{-0.02}$.

The disk-fed engine is in several respects a more comfortable solution than the isotropic one, and we regard it as the more physical of the two. The velocity dispersion and central density are lower by factors of a few and by more than two orders of magnitude respectively, which removes the most extreme feature of the isotropic solution. It is also the geometry that survives the compression criterion of Section~\ref{sec:compcond}: because flattening lowers $\sigma_{\rm coll}$ while leaving the clouds' capability $\sigma_{\rm comp} \simeq 271 \mhsix^{0.40}~{\rm km~s}^{-1}$ essentially unchanged, the requirement $\sigma_{\rm comp} > \sigma_{\rm coll}$ is met down to $M_{\rm h} = 6 \times 10^{5}~\msun$ at $f_{\ast} = 0.2$, against $8 \times 10^{6}~\msun$ for $f_{\ast} = 1$. The clouds are not merely consistent with the compressed cusp; they are capable of producing it. The clouds themselves become better analogues of CMZ objects, and the total gas mass again matches the observed molecular disks. Section~\ref{subsec:window} shows that $f_{\ast}$ is in fact not free at all, and that the fiducial value adopted here lies within the only range the model can consistently occupy.

The cost is the disruption rate. With $\Gamma_{\rm eq} = 1.1 \times 10^{-2}$~yr$^{-1}$ the enhancement over the canonical $10^{-4}$~yr$^{-1}$ is a factor of about a hundred rather than several thousand. We regard this as a more defensible claim in any case: it lies within the range of $10^{-3}$--$10^{-2}$~yr$^{-1}$ inferred for the galaxies that dominate the observed TDE population, rather than exceeding it. It is also comparable to the rates found for eccentric nuclear disks, where secular torques rather than momentum feedback drive stars to the loss cone \citep{Madigan:2018a,Wernke:2019a}; the two mechanisms are not exclusive, and a nucleus possessing both would plausibly reach higher rates than either alone.

Two conclusions of the isotropic model are weakened. The flattened cusp relaxes slowly, $\tau_{\rm relax,\ast} = 2.1 \times 10^{9} \mhsix^{3/2}$~yr, which is long compared to the engine lifetime; this is self-consistent, in that the anisotropy we assumed is not erased while the engine runs, but it also means the cusp retains whatever geometry it was built with and our treatment of it as a smooth flattened cusp may not be the optimal assumption. Second, the black hole's diet changes: $\langle \dot{M}_{\rm TDE}\rangle/\dot{M}_{\rm amb} = 0.25 f_{\rm B,-3}^{-1} \mhsix^{-0.96}$, so for a fiducial Bondi efficiency the ambient gas rather than the disrupted stars dominates the growth, and the ``black tide'' regime requires a less efficient accretor.

\begin{figure*}[t]
\centering\includegraphics[width=0.4\linewidth,clip=true]{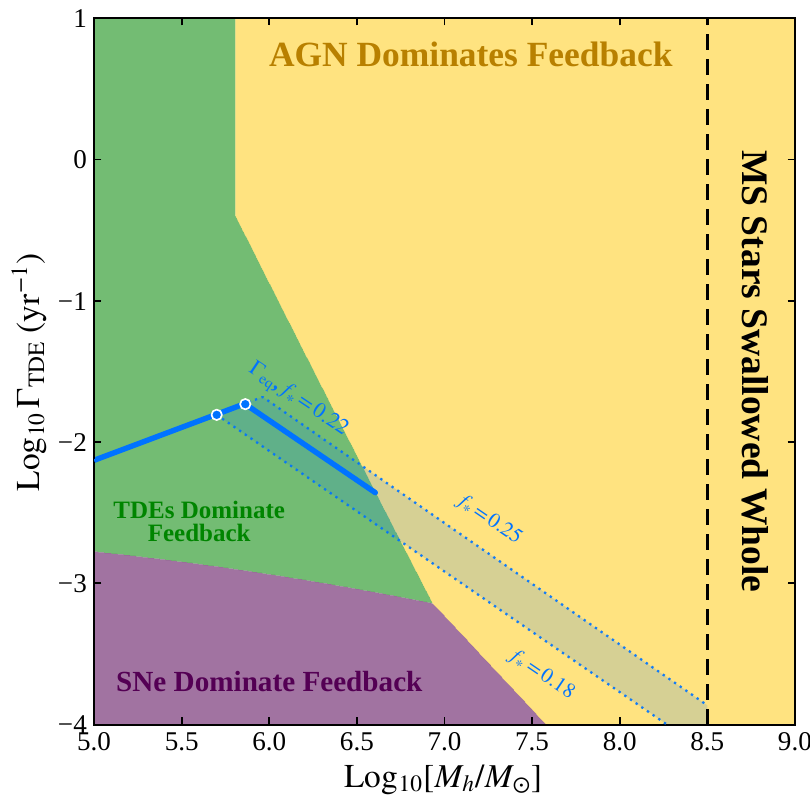}
\centering\includegraphics[width=0.4\linewidth,clip=true]{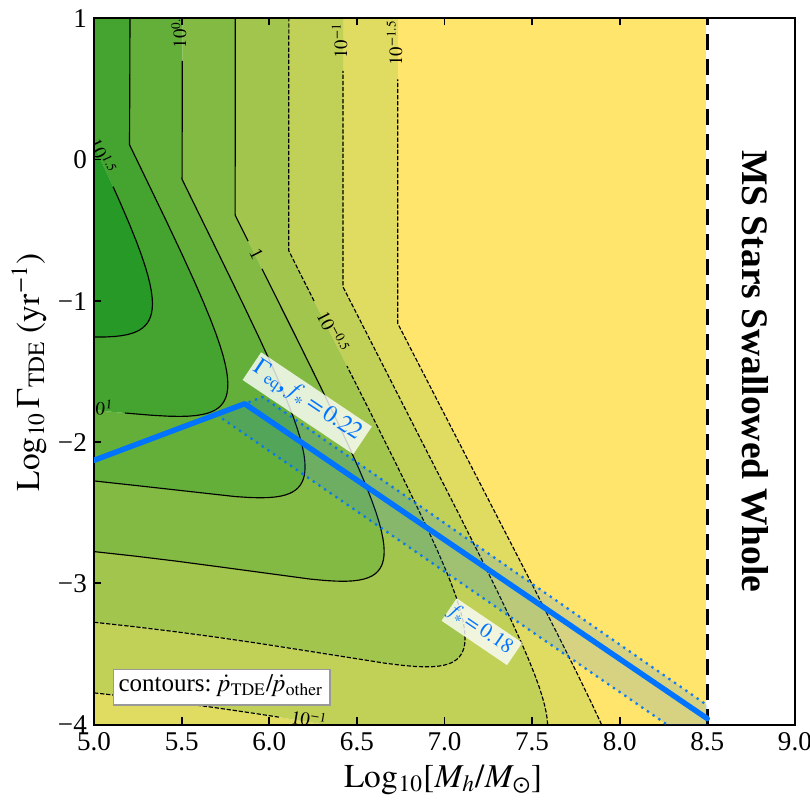}
\caption{Conditions under which tidal disruptions dominate all other sources of feedback in galactic nuclei. In the left-hand plot, the gold and green regions show where the AGN and TDE feedback mechanisms dominate for a given tidal disruption rate $\Gamma_{\rm TDE}$ and black hole mass $M_{\rm h}$. Within the region where TDEs dominate feedback, the equilibrium disruption rate relation is represented in blue. The shaded band spans the flattening allowed by Section~\ref{subsec:window}, bounded by the dotted curves at $f_{\ast} = 0.18$ and $f_{\ast} = 0.25$; the solid curve is the midpoint $f_{\ast} = 0.22$, drawn only where TDEs dominate the feedback budget. Each curve is the smaller of the collision-capped rate and the compression-limited rate of Section~\ref{sec:compcond}, and the filled circles mark where the two cross: below that mass the clouds rather than the collisions set the rate, and the curve turns over and rises with $M_{\rm h}$ as $\mhsix^{0.47}$. Note that both equilibrium curves leave the TDE-dominant region near $M_{\rm h} \simeq 5 \times 10^{6}~\msun$: the engine outweighs AGN feedback only at the low-mass end, which is the same restriction that Equation~(\ref{eq:fagn}) expresses in terms of $f_{\ast}$. The right hand plot shows the ratio of momentum injection from TDEs $\dot{p}_{\rm TDE}$ to that from all other momentum injection processes, $\dot{p}_{\rm other}$. If a system occupies any position within the green region, it will be driven toward the equilibrium line by adjusting $\Gamma_{\rm TDE}$.}
\label{fig:dominance}
\end{figure*}

\section{Starting and Stopping the Engine}\label{sec:spark}

As shown in Figure \ref{fig:dominance}, the minimum required disruption rate to sustain a disruption engine is $\gtrsim 10^{-4}$~yr$^{-1}$, and for some galaxies this rate might already be realized just from two-body relaxation \citep{Stone:2016a}. However, because of the dependence of the TDE momentum feedback on $M_{\rm h}$ (Equation (\ref{eq:pudr})), the $10^{-4}$~yr$^{-1}$ requirement is only sufficient for the most-massive black holes, $M_{\rm h} \gtrsim 10^{8} M_{\odot}$, such black holes tend to have longer relaxation times and thus lower rates of tidal disruption. For lower black hole masses the disruption rate requirements are stiffer, with $\gtrsim 10^{-3}$~yr$^{-1}$ likely being required for $10^{6} M_{\odot}$ black holes.

Once established, a disruption engine will maintain a high equilibrium rate so long as the galactic nucleus continues to be surrounded by molecular gas, and will naturally inhibit the local SNe rate and set the level of AGN activity. Thus, all that is required is to temporarily raise the disruption rate (``spark'' the engine) above the critical threshold necessary to enter into the disruption engine state. So long as this enhancement lasts for longer than the time it takes for massive stars to form and produce SNe ($\sim 5 \times 10^{7}$~yr$^{-1}$), the disruption engine will continue to operate even after the mechanism for the temporary increase stops operating.

A natural mechanism for temporarily enhancing the tidal disruption rate for a few $10^{8}$~yr is the merger of the black hole at the center of the nuclear cluster with a secondary black hole of comparable mass. Some models have predicted that such mergers can yield extremely high disruption rates \citep{Chen:2009a}, but even modest enhancements in the rate by a factor of $\sim 10$ above the nominal value can mean the difference between being in the self-sustaining engine state or not.

While it might seem natural to ascribe the rate increase predicted as the result of mergers with the observed rate enhancement, the primary issue is that the merger will only result in the disruption of a fixed number of stars that will be significantly less than the mass of the secondary black hole. The faster the merger, the higher the disruption rate, but at the expense of a shorter duration of enhancement. \citet{Wegg:2011a} estimate that only 3\% of all stellar disruptions come as the result of black hole mergers, a number that would need to be inverted (a few percent of disruptions around {\it single} black holes) to explain the observed enhancement.

If a black hole merger is responsible for sparking the engine, a plausible sequence of events that lead to a black hole being in the disruption engine state are:

\begin{enumerate}
\item Two galaxies merge, driving gas to the center of the more-massive galaxy. This yields a burst of star formation.
\item The two central black holes merge after 1 Gyr, enhancing the TDE rate. TDEs have the advantage that they accrete at rates much greater than Eddington, converting a larger fraction of the emergent energy into wide relativistic outflows.
\item The injection of kinetic energy suppresses star formation by sustaining turbulent support in the molecular clouds, in the same manner inferred for the Milky Way's Central Molecular Zone \citep{Longmore:2013a,Kruijssen:2014a}.
\item The clouds settle into a belt close enough to act as massive perturbers, moving the nucleus from the empty toward the full loss cone regime \citep{Spitzer:1951a,Perets:2007a,Merritt:2013a}.
\item The enhanced rate of TDEs maintains the suppressed star formation despite the massive gas reservoir. Enhanced TDEs continued until the gas reservoir is exhausted or ejected from the galaxy as a molecular outflow.
\end{enumerate}

\subsection{The Flattening is Determined, Not Assumed}\label{subsec:window}

\begin{figure*}[t]
\centering
\includegraphics[width=\textwidth]{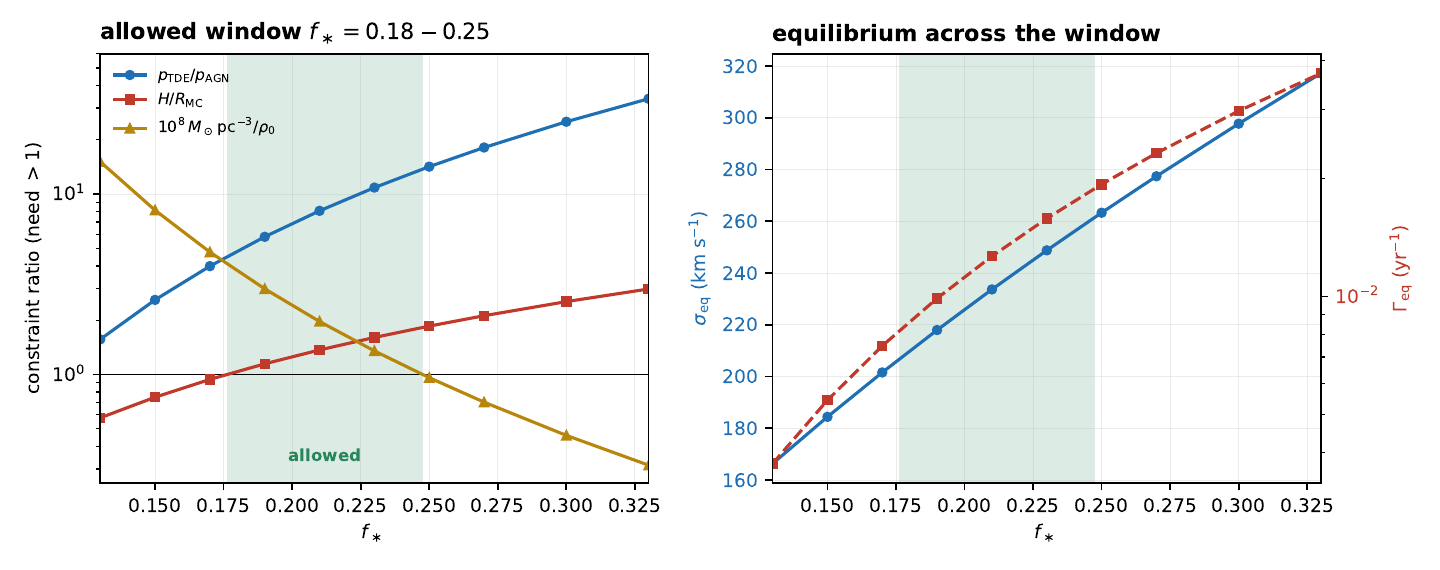}
\caption{{\it Left:} the three $f_{\ast}$-dependent requirements, each plotted as a ratio that must exceed unity. TDE momentum dominance and the requirement that clouds fit within the disk both fail for flat systems; the central density of the cusp rises above the densest observed nuclear star clusters, $10^{8}~\msun~{\rm pc}^{-3}$, for round ones. The shaded band is the allowed window, $0.18 \le f_{\ast} \le 0.25$. {\it Right:} the equilibrium dispersion and disruption rate across the same range.}
\label{fig:window}
\end{figure*}

Sections~\ref{subsec:engine-state} and \ref{subsec:disk} treat the cusp flattening $f_{\ast}$ as a parameter to be chosen. It need not be. Three of the requirements the engine must satisfy depend on $f_{\ast}$, and they do not all pull in the same direction; taken together they close a window.

The first two push toward rounder systems. Flattening suppresses the collision rate by $f_{\ast}$, which lowers $\sigma_{\rm coll}$ and hence $\Gamma_{\rm eq}$; but the ambient density, and with it the Bondi rate that powers the AGN, falls more slowly. The margin by which TDE momentum beats AGN radiation pressure therefore shrinks as the system is flattened, and the premise of Section~\ref{sec:engine} fails below
\begin{equation}
f_{\ast} \gtrsim 0.14 \quad\quad (p_{\rm TDE} = p_{\rm AGN}).\label{eq:fagn}
\end{equation}
Separately, the clouds must fit inside the disk that contains them. The belt solve returns $R_{\rm MC}$ independently of the disk half-thickness $H = f_{\ast} r_{\rm b}$, and for a thin enough disk the derived clouds are thicker than the structure they are supposed to occupy. Requiring $H > R_{\rm MC}$ gives
\begin{equation}
f_{\ast} \gtrsim 0.18 \quad\quad (H = R_{\rm MC}),\label{eq:fgeo}
\end{equation}
which is the more restrictive of the two.

The third requirement pushes the other way, and it is an observational ceiling rather than an internal one. Because $\sigma_{\rm eq} \simeq 6.5 \times 10^{2} f_{\ast}^{0.66}~{\rm km~s}^{-1}$ rises with $f_{\ast}$ while $a_{\rm h} \propto \sigma^{-2}$ falls, the central density of the cusp climbs very steeply as the system is made rounder,
\begin{equation}
\rho_{0} \simeq 2.4 \times 10^{10}\, f_{\ast}^{3.96}\, \mhsix^{-1.8}~\msun~{\rm pc}^{-3}.\label{eq:rho0f}
\end{equation}
At $f_{\ast} = 1$ this is $2 \times 10^{10}~\msun~{\rm pc}^{-3}$ for a $10^{6}~\msun$ hole, exceeding the densest observed nuclear star clusters by more than two orders of magnitude \citep{Neumayer:2020a}. Requiring instead that the cusp be no denser than the densest nuclei actually observed, $\rho_{0} \lesssim 10^{8}~\msun~{\rm pc}^{-3}$, gives
\begin{equation}
f_{\ast} \lesssim 0.25 \quad\quad (\rho_{0} = 10^{8}~\msun~{\rm pc}^{-3}).\label{eq:ffloor}
\end{equation}

Equations~(\ref{eq:fgeo}) and (\ref{eq:ffloor}) bracket the allowed range,
\begin{equation}
0.18 \lesssim f_{\ast} \lesssim 0.25,\label{eq:fwindow}
\end{equation}
shown in Figure~\ref{fig:window}. What matters is not that the band is narrow but that it is bounded at all: the model cannot be made arbitrarily flat without losing TDE momentum dominance and the geometric coherence of its own clouds, nor arbitrarily round without demanding a cusp denser than any that has been observed. The flattening was introduced in Section~\ref{subsec:disk} as a fiducial choice, set to $f_{\ast} = 0.2$ by analogy with observed circumnuclear disks; it falls near the middle of the range those requirements allow. The requirement is testable: the engine needs a nuclear stellar structure with an axis ratio near $1{:}5$, not a sphere and not a thin disk.

At the midpoint of the window, $f_{\ast} = 0.22$ and $M_{\rm h} = 10^{6}~\msun$, the engine has
\begin{equation}
\begin{split}
\sigma_{\rm eq} &= 2.4 \times 10^{2}~\mhsix^{0.03}~{\rm km~s}^{-1}, \\
\Gamma_{\rm eq} &= 1.4 \times 10^{-2}~\mhsix^{-0.84}~{\rm yr}^{-1},
\end{split}\label{eq:eqwindow}
\end{equation}
with $r_{\rm b} = 4.8$~pc, $R_{\rm MC} = 0.71$~pc, ${\cal M} = 32$, $n_{\rm MC} = 1.9\times10^{5}$~cm$^{-3}$, $M_{\rm gas} = 3.9\times10^{5}~\msun$, $A_{V} = 50$, $p_{\rm TDE}/p_{\rm AGN} = 9.4$, and a mass floor of $7.3 \times 10^{5}~\msun$. The central density is $\rho_{0} = 6 \times 10^{7}~\msun~{\rm pc}^{-3}$, high but no longer beyond the densest observed nuclear star clusters, and a factor of $\sim 300$ below the isotropic solution.

Two features of the window deserve emphasis. First, the extinction is invariant across it: $A_{V}$ varies only between $49.5$ and $50.2$ over the whole allowed range, and indeed only between $47$ and $53$ over the far wider range $0.06 \le f_{\ast} \le 1$. The obscuration prediction of Section~\ref{subsec:obscured} is therefore independent of the geometry, and is the most robust observational statement the model makes. Second, the mass dependence of the rate is steep, $\Gamma_{\rm eq} \propto M_{\rm h}^{-0.84}$, and this exponent is stable across the window and across the collision normalization. It is the normalization, not the slope, that carries the uncertainty.

The three bounds are not equally secure. Equation~(\ref{eq:fgeo}), the binding lower bound, is purely geometric and depends on nothing else. Equation~(\ref{eq:fagn}) depends on the assumed Bondi efficiency, and a less efficient accretor ($f_{\rm B} < 10^{-3}$) relaxes it. Equation~(\ref{eq:ffloor}) is an observational ceiling rather than an internal inconsistency, and it inherits the mass dependence of Equation~(\ref{eq:rho0f}): the bound scales as $f_{\ast} \lesssim 0.25 (M_{\rm h}/10^{6}~\msun)^{0.45}$, so it is most restrictive for the low-mass holes that dominate the observed TDE host population and weakens for more massive ones. We also note that the compression criterion of Section~\ref{sec:compcond} supplies a second, weaker upper bound of the same kind: above $f_{\ast} \simeq 0.5$ the cusp is consumed faster than two-body relaxation can rebuild it, $N_{\ast}/\Gamma_{\rm eq} < \tau_{\rm relax,\ast}$, and the relaxed Bahcall-Wolf profile assumed throughout is no longer self-consistent. Since Equation~(\ref{eq:ffloor}) is the more restrictive, we quote it alone.

\section{Discussion}\label{sec:discussion}

\subsection{The lifetime of an engine}\label{subsec:lifetime}

The engine is not perpetual: it consumes both the stars of the cusp and the gas of the disk. Disruptions remove stellar mass at a rate $m_{\ast}\Gamma_{\rm eq}$, so the cusp is exhausted in
\begin{equation}
t_{\rm cusp} = \frac{N_{\ast}}{\Gamma_{\rm eq}} = 2.8 \times 10^{8} \mhsix^{1.84}~{\rm yr},
\end{equation}
while the disk, if it resupplies the cusp through star formation, is drained in
\begin{equation}
t_{\rm gas} = \frac{M_{\rm disk}}{m_{\ast} \Gamma_{\rm eq}} = 5.5 \times 10^{7} \mhsix^{2.31}~{\rm yr},
\end{equation}
both evaluated at the window midpoint $f_{\ast} = 0.22$ of Section~\ref{subsec:window}. Gas exhaustion is the binding constraint, and it sets an engine lifetime of a few $10^{8}$~yr for a $10^{6}~\msun$ black hole, provided the disk is not replenished from larger radii. Since the circumnuclear reservoirs of post-starburst galaxies reach $\sim 10^{9}~\msun$, three orders of magnitude above the disk mass required here, resupply from outside is likely to extend this, and the cusp consumption time then becomes the ceiling.

We note that these two timescales are close to one another in the isotropic solution and separate by a factor of a few in the disk-fed one; their ratio is a diagnostic of the geometry rather than a robust prediction, and we caution against reading physical significance into the near-coincidence of the isotropic case. What is robust is the order of magnitude: engines live for $10^{8}$--$10^{9}$~yr, neither so briefly that they would be vanishingly rare nor so long that they would be common.

This overlaps the ages at which post-starburst spectral signatures are strongest, and TDE hosts are drawn heavily from that population \citep{French:2020a,Hammerstein:2023a}. The delay time distribution measured by \citet{Shepherd:2026a} sharpens the comparison. Constructing the TDE rate per galaxy as a function of time since the burst, normalized by a control sample matched in the same way, they find that the rate {\it rises} with post-burst age and peaks near $1$~Gyr in the subsample of hosts with a significant measured burst. They note that the models they compare --- overdense nuclei, radial anisotropy, black hole binaries, AGN disks --- predict rates that decline with post-burst age, and are in tension with the data for that reason. The engine makes two predictions that bear on this, and they are not independent of one another.

The first is a delay. The compression criterion of Section~\ref{sec:compcond} is a threshold in black hole mass, not in time: below $M_{\rm floor} \simeq 7 \times 10^{5}~\msun$ at $f_{\ast} = 0.22$ the clouds cannot compress the cusp and no engine exists. A nucleus whose black hole is seeded below that threshold produces no enhanced disruption rate at the moment of the burst, however much gas it acquires. It does, however, accrete: growing at a fraction $f_{\rm Edd}$ of the Eddington rate with radiative efficiency $\epsilon$, the hole crosses the threshold after
\begin{equation}
\begin{split}
t_{\rm on} &= \frac{\epsilon \sigma_{\rm T} c}{4 \pi G m_{\rm p} f_{\rm Edd}}
   \ln\!\left(\frac{M_{\rm floor}}{M_{\rm i}}\right) \\
&\simeq 8.8 \times 10^{8} \left(\frac{f_{\rm Edd}}{0.05}\right)^{-1}
   \left(\frac{\epsilon}{0.05}\right) \\
&\quad\quad \times \frac{\ln\left(M_{\rm floor}/M_{\rm i}\right)}{1.99}~{\rm yr},
\end{split}
\label{eq:ton}
\end{equation}
normalized to a seed of $M_{\rm i} = 10^{5}~\msun$ and to the Eddington ratio $f_{\rm Edd} \simeq 0.05 f_{\rm B,-3}$ that the equilibrium itself supplies. Seeds of $5 \times 10^{4}$, $10^{5}$, $2 \times 10^{5}$ and $4 \times 10^{5}~\msun$ switch on at $1.2$~Gyr, $8.8 \times 10^{8}$, $5.7 \times 10^{8}$ and $2.7 \times 10^{8}$~yr respectively, and direct integration of the coupled growth and gas-depletion equations reproduces Equation~(\ref{eq:ton}) to better than $2\%$. Because \citet{Shepherd:2026a} normalize to a control sample, the quantity to compare with is the mean rate per galaxy, which is the fraction of nuclei hosting an active engine multiplied by the rate per engine. Engines seeded above the threshold run immediately and switch off as their gas is consumed; engines seeded below it are silent until Equation~(\ref{eq:ton}) brings them across. The active fraction therefore rises while seeds are still crossing and falls once gas exhaustion overtakes the supply of new ones. Integrating over seeds distributed as ${\rm d}N/{\rm d}\log M \propto M^{\alpha}$, the mean rate per galaxy peaks between $1.1$ and $1.7$~Gyr for $-0.5 \le \alpha \le -0.3$, the range implied by the faint-end slope of the galaxy stellar mass function. The location of the peak is set by Equation~(\ref{eq:ton}) and is insensitive to $\alpha$, to the upper limit of the seed distribution, and to whether ambient accretion continues after switch-on; the amplitude of the rise, a factor of $1.1$--$3.7$ across those choices, is not.

The second prediction concerns mass, and we regard it as the more discriminating of the two. Engines that have just crossed the threshold sit, by construction, at $M_{\rm h} \simeq M_{\rm floor} \simeq 7 \times 10^{5}~\msun$, and because $\Gamma_{\rm eq} \propto M_{\rm h}^{-0.84}$ they carry the highest disruption rate in the population, $\Gamma_{\rm eq} \simeq 1.9 \times 10^{-2}$~yr$^{-1}$. The late-time enhancement is therefore contributed by black holes near the bottom of the engine's mass range rather than the top, and the engine hypothesis predicts that the hosts responsible for the peak should have {\it low} black hole masses, of order $10^{6}~\msun$. This is the opposite of what the two mechanisms best able to produce a late peak predict. As \citet{Shepherd:2026a} discuss, the black hole binary models of \citet{Melchor:2024a} place their peak at comparable ages once shifted by the coalescence time, but \citet{Melchor:2025a} find that the binary disruption rate rises with the mass of the disrupting hole up to $10^{8}~\msun$; radial anisotropy models likewise favor high masses, because radial biases relax away faster in low-mass nuclei. \citet{Shepherd:2026a} test exactly this. They find no preference for high black hole mass among either the high-burst-fraction or the post-starburst hosts, find that post-starburst hosts occupy the {\it lower} end of the stellar mass range of the parent TDE sample, and find that low-stellar-mass hosts show a higher rate at $1$~Gyr than high-mass ones. They use these results to disfavor the binary and anisotropy explanations. The same measurements are what the engine predicts.

Several things temper this. The rise is measured in the subsample of $15$ hosts with a significant burst; the post-starburst subsample, normalized to its own control, is flatter and its burst ages are not statistically distinguishable from the control. The peak is driven in part by the scarcity of control galaxies with bursts that old rather than by a large number of TDE hosts. What the model supplies, and what we think is worth testing, is a joint prediction: the peak of the delay time distribution should track the time for a seed black hole to grow into the engine regime, should shift to earlier times in more efficiently accreting nuclei, and should be carried by hosts with black hole masses near $10^{6}~\msun$ rather than above it.

\subsection{An obscured population of disruptions}\label{subsec:obscured}

The equilibrium ambient density implies a visual extinction $A_{\rm V} \simeq 50$ along a sightline that passes through the gas. A disruption seen through the clouds is therefore not a source that optical surveys can find: it is reprocessed by dust and emerges in the infrared.

The obscuration is not isotropic, however, and this is where the disk geometry of Section~\ref{subsec:disk} makes a sharper prediction than the isotropic belt. If the gas subtends only a fraction $f_{\Omega}$ of $4\pi$, then only that fraction of sightlines is extinguished; a nucleus viewed close to face-on offers an unobstructed view of the black hole even while an edge-on view is buried under $A_{\rm V} \simeq 50$. This is precisely the orientation dependence that underpins the unified model of active nuclei, in which the same object appears obscured or unobscured according to whether the torus intercepts the line of sight \citep{Netzer:2015a,RamosAlmeida:2017a}.

The engine hypothesis therefore does not predict that all high-rate nuclei are hidden. It predicts that the obscured fraction of disruptions {\it equals the covering factor of the gas}, which for our fiducial $f_{\Omega} = 0.2$ means that roughly one engine disruption in five is lost to dust while the remainder are seen at close to their intrinsic brightness. This is a considerably more falsifiable statement than the isotropic case, in which every sightline is blocked and the engine population would be invisible in the optical altogether. It also removes an obvious objection: were engines uniformly obscured, they could not contribute to the optically selected samples in which the post-starburst preference was found in the first place.

This is a testable statement, and the evidence has moved in its favor since this work was begun. Mid-infrared searches have uncovered a population of dust-obscured disruptions whose hosts are gas-rich and star-forming, and which are largely invisible to optical surveys \citep{Masterson:2024a}. The volumetric rates inferred from optical surveys \citep{vanVelzen:2021a,Yao:2023a} fall short of loss-cone predictions by roughly an order of magnitude, which is the direction and approximate magnitude that a substantially obscured high-rate population would produce. We emphasize that the engine hypothesis does not require every obscured TDE to arise in an engine; it requires that nuclei with the highest rates be preferentially, but not exclusively, obscured. Measuring the ratio of infrared-selected to optically selected disruptions among post-starburst hosts would in principle measure $f_{\Omega}$ directly.

\subsection{Rapid black hole growth through tidal disruptions}\label{subsec:growth}

For a Bondi fraction $f_{\rm B} = 10^{-3}$, Equation (\ref{eq:eqaccratio}) predicts that the rate at which matter is accreted by the black hole from tidal disruptions exceeds the rate at which it is acquired from the ambient medium by a factor of a few for $M_{\rm h} \sim 10^{6}~\msun$, placing the engine in the ``black tide'' mode of growth \citep{Young:1977a}. This conclusion is geometry-dependent: in the disk-fed engine of Section~\ref{subsec:disk} the same ratio is $0.25 f_{\rm B,-3}^{-1}$, and the ambient gas instead dominates unless the black hole is a relatively inefficient accretor. This ratio is entirely dependent on how quickly the black hole accretes gas from the ambient medium; in cases where the accretion rate is closer to 1\% of Bondi \citep[e.g. M87,][]{Kuo:2014a}, the black hole's growth is still dominated by the accretion of the surrounding gas, whereas a Bondi fraction of $\sim 10^{-4}$ \citep[e.g. the MW,][]{Quataert:2000a} would place growth firmly in the tidal channel. Because the accreted mass is delivered in discrete, super-Eddington episodes rather than continuously, the resulting duty cycle is highly intermittent, comparable to the flickering inferred for low-luminosity AGN \citep{Schawinski:2015a}.

Over an engine lifetime the black hole gains of order its own mass in stellar material in either geometry, since the lower disruption rate of the disk-fed solution is compensated by its longer lifetime. Engines are therefore not a negligible contribution to the growth of black holes in this mass range, though they are too rare to compete with radiatively efficient accretion in the global budget.

\subsection{Suppressed star formation and the CMZ analogy}\label{subsec:sf}

The clouds are turbulently supported at ${\cal M} \simeq 40$ and sit at densities of $10^{4}$--$10^{5}~{\rm cm}^{-3}$ while forming few stars. This combination is unusual but not unprecedented: it is the condition of the Milky Way's Central Molecular Zone, where the star formation rate falls an order of magnitude below what the dense gas mass would predict \citep{Longmore:2013a,Kruijssen:2014a,Barnes:2017a}. In the CMZ this suppression is generally attributed to the elevated turbulent pressure raising the density threshold for collapse; in an engine, the same suppression is maintained by UDR driving, and it is self-sustaining rather than incidental.

The observational corollary is that engines should appear as nuclei rich in dense-gas tracers (HCN, HCO$^{+}$) but weak in star formation indicators, alongside a low level of AGN activity, $L_{\rm AGN}/L_{\rm Edd} \sim 0.06 f_{\rm B,-3}$. Such centrally concentrated, weakly star-forming molecular reservoirs are routinely observed in post-starburst galaxies \citep{French:2015a,French:2018a,Smercina:2018a}, which are the same hosts in which TDEs are over-represented, and are seen in detail in NGC~1266 \citep{Alatalo:2011a,Alatalo:2014a,Alatalo:2015a}.

\subsection{Caveats}\label{subsec:caveats}

The treatment here is deliberately an order-of-magnitude one, and several of its simplifications would need to be relaxed before the numbers above could be taken as predictions rather than as scalings.

The first concerns the loss cone. The equilibrium cusp is not in either limiting regime: at the sphere of influence the loss cone is full, with $q_{\ast} = 1.7 \mhsix^{-0.27}$ for the isotropic solution and $q_{\ast} = 12 \mhsix^{-0.33}$ at $f_{\ast} = 0.22$, while $q \propto r^{9/4}$ carries it into the diffusive regime further in. The critical radius therefore lies inside the cusp rather than outside it, at $r_{\rm crit} \simeq 0.8 a_{\rm h}$ in the isotropic case and $\simeq 0.3 a_{\rm h}$ in the disk-fed one, and both regimes contribute to the rate. This is the situation Equation~(\ref{eq:gtde}) is constructed to handle, and it is why we use the bridged flux rather than either limiting form: a pure pinhole treatment would not only misestimate the rate but, as noted in Section~\ref{subsec:conditions}, would remove the $\sigma$-dependence that makes the equilibrium exist at all. The residual uncertainty in $\Gamma_{\rm eq}$ is therefore not the loss-cone regime but the collision normalization $\Lambda_{\rm c}$, which spans two orders of magnitude in the rate for $1 \le \Lambda_{\rm c} \le 10$ while leaving the mass exponent stable to $\pm 0.1$. A direct $N$-body or Fokker-Planck treatment of a flattened cusp embedded in a massive-perturber disk would tighten this, and would also test our assumption that the disruption rate is insensitive to flattening while the collision rate is not; that asymmetry is what sets the scaling in Equation~(\ref{eq:gammaeqf}), and it deserves a more careful treatment than the scaling argument we have given it.

The compression criterion of Section~\ref{sec:compcond} carries a limitation of its own. Equation~(\ref{eq:cond4}) compares a relaxation time driven by the clouds with one driven by the stars, but the two are not evaluated at the same place: the solution puts the clouds at $r_{\rm b}$ and the stars within $a_{\rm h}$. Treating the belt as a shell fixed at $r_{\rm b}$ is, however, the least favorable reading of its own geometry. Equation~(\ref{eq:rhomc}) places the clouds at the margin of tidal stability, so they are not static: on eccentric orbits they reach well inside $r_{\rm b}$, where the tidal field exceeds their self-gravity and shears them, and the resulting streams and fragments carry cloud material inward on a continuum of radii rather than leaving it confined to the belt. The perturbers that torque the cusp are therefore this sheared, inward-extending debris rather than a smooth shell held at a single radius, and evaluating $t_{\rm relax,MC}$ with the cloud density is the appropriate choice for it.

A choice is embedded in Equation~(\ref{eq:cond4}) that should be made explicit. The clouds fill only $\sim 20\%$ of the belt by volume, so the internal cloud density and the density smoothed over the belt differ by a factor of about five: the smoothed value would raise the mass floor by roughly a factor of two and narrow the window of Section~\ref{subsec:window} to a point near $f_{\ast} \simeq 0.19$. We use the internal density. The relevant physics is the torquing of cusp stars by discrete massive perturbers during close passages, not the mean field of a smooth shell, and the molecular gas in galactic nuclei is strongly clumped rather than smoothly distributed: the Milky Way's own circumnuclear disk resolves into distinct clouds with a low volume filling factor \citep{Christopher:2005a,MonteroCastano:2009a}, and the same is true of the CMZ clouds on which our cloud model is based \citep{Longmore:2013a,Kruijssen:2014a}. In a clumpy medium the encounter rate with individual clouds, and hence the angular momentum diffusion they drive, is set by the density within the clumps. We note the alternative because the window is narrow enough that the choice is not cosmetic, but we regard the clumped interpretation as the physical one.

What would settle both points is a direct calculation of the angular momentum transferred to cusp stars by a population of tidally shearing clumps on eccentric orbits, tracking how far inward the debris penetrates and how much of it survives as discrete perturbers. Until that exists, the positive branch of the loop rests on a plausibility argument rather than a computed torque.

We have also treated the clouds as a single population of identical objects at a single radius, ignored the dynamical friction that would cause them to spiral inward over $\sim 10^{9}$~yr, and adopted a single aspect ratio for both the gas and the stars, whereas a real nucleus would present a range of inclinations and a warped, possibly eccentric, disk. Each of these is a factor-of-a-few effect on quantities that already carry factor-of-a-few uncertainties, and none of them changes the central result: that a self-consistent, self-sustaining solution exists, that it lies a few hundred times above the canonical disruption rate, and that it is obscured.

\section{Implications for Observed Populations}\label{sec:implications}

Two populations of galaxies are close enough to the engine state described above that the model makes contact with them directly. We take each in turn.

\subsection{Enhancement of k+A galaxies in clusters}\label{subsec:ka}

k+A galaxies show central molecular regions with a large amount of mass, $\sim 10^{9}~\msun$, molecular outflows, and suppressed star formation \citep{French:2015a,Smercina:2018a}, a combination exhibited in detail by NGC~1266 \citep{Alatalo:2011a,Alatalo:2014a,Alatalo:2015a}. The disk mass required by the engine, $\sim 7 \times 10^{5} \mhsix^{1.14}~\msun$, is a small fraction of these reservoirs, so a k+A nucleus has ample fuel to sustain an engine for the $\sim 10^{8}$~yr derived above. The engine hypothesis therefore does not require these galaxies to be unusual in their total gas content, only in the configuration of the innermost few parsecs of it.

Three properties of the k+A population make them natural hosts. First, their morphologies are disturbed: a large fraction show tidal features and other signatures of a recent major merger \citep{Zabludoff:1996a}, which supplies both the violent relaxation needed to spark the engine (Section~\ref{sec:spark}) and the inflow of gas needed to build the disk. Second, and less obviously, they retain that gas. Post-starburst galaxies were long assumed to have exhausted or expelled their molecular reservoirs, but sensitive CO and dust measurements find substantial cold gas surviving for hundreds of Myr after the burst \citep{French:2015a,Smercina:2018a}. Third, that surviving gas is not forming stars at the rate its density would imply \citep{French:2018a}, which is the same decoupling of dense gas from star formation that the engine requires and that the CMZ exhibits.

Taken together, these are the conditions the model needs, and they are observed to persist on precisely the timescale over which we find an engine can run. However, a merger that sparks an engine also produces a k+A spectrum and a central gas reservoir independently, so the association of TDEs with k+A hosts is equally consistent with the engine being a consequence of the same event rather than a cause of the elevated rate. Distinguishing the two requires the nuclear observations described above, not host-galaxy demographics.

\subsection{Connection to Little Red Dots (LRDs)}\label{subsec:lrd}

The compact, red, high-redshift sources uncovered by {\it JWST} and known as little red dots have been proposed to be powered by tidal disruptions. \citet{Bellovary:2025a} suggested that LRDs are recurring TDEs in clusters that have undergone runaway collapse to an intermediate-mass black hole, a system that would be compact, UV-emitting, and display the broad H$\alpha$ emission that characterizes the population. \citet{Perger:2025a} subsequently showed that radio stacking limits on LRDs are compatible with radio-quiet TDEs as the underlying process. Independently, \citet{Pacucci:2025a} find that a purely stellar reading of the LRD photometry implies central densities of $10^{4}$--$10^{5}~\msun~{\rm pc}^{-3}$, and in the most extreme cases up to $\sim 10^{9}~\msun~{\rm pc}^{-3}$, and conclude that the dense residual stellar core left behind by seed formation can sustain high rates of tidal disruption. The abundance, compactness and redshift distribution of the population are separately reproduced if LRDs occupy the low-spin tail of the halo distribution \citep{Pacucci:2025b}.

What this literature does not supply is a mechanism that sets the disruption rate. \citet{Bellovary:2025a} notes that $\Gamma_{\rm TDE} \sim 10^{-4}$~yr$^{-1}$ suffices to reconcile the observed LRD number density with the predicted abundance of runaway-collapse clusters, since with flare durations of order a year that rate is also the duty cycle. This is a population-level requirement rather than a prediction for any individual system, and the cluster-dynamics models they compare in fact deliver considerably more. Both the runaway tidal-encounter calculation of \citet{Stone:2017a} and the $N$-body models of \citet{Rizzuto:2023a} give per-system rates that climb steeply with black hole mass, reaching $10^{-2}$--$10^{-1}$~yr$^{-1}$ by $M_{\rm h} = 10^{6}~\msun$ for stellar densities of $10^{8}~\msun~{\rm pc}^{-3}$. What is missing is not a high enough rate but a reason for the nucleus to sit at the density that produces one. Our equilibrium is precisely such a mechanism, and the densities it produces are of the same order as those inferred for LRD cores: the cusp normalization of Section~\ref{subsec:engine-state} corresponds to $\sim 10^{9}~\msun~{\rm pc}^{-3}$ in the isotropic case and $\sim 10^{7}~\msun~{\rm pc}^{-3}$ in the disk-fed case. We caution that this is not a like-for-like comparison, since the LRD densities are inferred within effective radii of $\sim 100$~pc while ours are cusp normalizations at $a_{\rm h} \sim 0.05$--$0.3$~pc.

Our equilibrium rates across the allowed window, $8.6 \times 10^{-3}$--$2.1 \times 10^{-2}$~yr$^{-1}$, fall squarely inside the range those models reach at the upper end of the mass distribution, and the stellar density we require, $10^{7}$--$10^{9}~\msun~{\rm pc}^{-3}$, brackets the $10^{8}~\msun~{\rm pc}^{-3}$ their fiducial curves assume. Because a higher rate per system implies a higher duty cycle, fewer hosts are needed to reproduce the observed abundance, which would ease the tension \citet{Bellovary:2025a} identifies between the required host density and the $\Lambda$CDM halo mass function, and weaken the case for the top-heavy initial mass function invoked to resolve it.

We would nonetheless caution against reading too much into the numerical agreement, because the two calculations hold different quantities fixed. The cluster models adopt a velocity dispersion and stellar density as inputs and obtain a rate that rises with black hole mass; in our equilibrium these are outputs, determined self-consistently, and the resulting rate falls steeply with mass as $M_{\rm h}^{-0.84}$. The velocity dispersions also differ, ours being several times larger than the $40~{\rm km~s}^{-1}$ of their fiducial models, with the higher stellar density compensating. Ours moreover describes a momentum-driven equilibrium around a $10^{5}$--$10^{8}~\msun$ black hole already embedded in molecular gas, while the LRD systems are $\sim 10^{4}~\msun$ seeds newly assembled by stellar collisions. That two such different routes arrive at comparable rates is encouraging, but it may be coincidence.

The role of stellar collisions is in fact inverted between the two pictures: in the LRD seed-formation scenario, collisions are the engine; runaway mergers in a dense cluster assemble a very massive star that contracts and collapses to a black hole of $\sim 10^{4}~\msun$ \citep{Pacucci:2025a}. In our model, collisions are the brake, scattering stars out of the loss cone and capping the disruption rate through the condition of Section~\ref{sec:collcond}. Both processes become important at comparable stellar densities, and Equation~(\ref{eq:cond3}) is in effect the boundary between the two regimes. A nucleus that has just finished building a seed by runaway collision should sit close to that boundary, which offers a natural way to join the two descriptions: the collisional phase that makes the black hole gives way to a collisionally regulated phase that feeds it.

Two obstacles stand in the way of taking the connection further. First is metallicity: our equilibrium relies on CO line cooling and on dust to hold the clouds near $10$~K; in the metal-poor environments where runaway collapse is expected to operate at $z \sim 5$, neither is available, the clouds cannot reach such low temperatures, and the second of our conditions would have to be rederived with molecular hydrogen or atomic cooling in place of Equation~(\ref{eq:lmc}). The engine as formulated here is a metal-enriched, low-redshift mechanism. The second is obscuration: we predict $A_{\rm V} \simeq 50$ through the gas, which must be reprocessed into the infrared, whereas LRDs show a marked deficit of both hot and cold dust emission \citep{Setton:2025a,Chen:2025a}. This is only reconcilable if the clouds are arranged in a disk rather than a spherical shell, so that the covering factor is $f_{\Omega} \sim 0.2$ and the reprocessed luminosity is suppressed by the same factor, with most sightlines unobscured. The isotropic geometry of Section~\ref{subsec:engine-state} is straightforwardly ruled out by the LRD dust constraints; only the disk-fed configuration of Section~\ref{subsec:disk} survives them. We note in passing that if the reddening of LRDs arises from a dense gas envelope rather than from dust, as in the ``black hole star'' interpretation of \citet{Naidu:2025a}, then a gas-enshrouded nucleus of the kind our engine constructs is qualitatively the right picture regardless.

\bigskip
\acknowledgements
\begin{sloppypar}
We thank Fabio~Antonini, Blakesley~Burkhart, Xian~Chen, Charlie~Conroy, Robert~Fisher, Nathan~Goldbaum, Luke~Zoltan-Kelley, Morgan~MacLeod, Peter~Maksym, Ilya~Mandel, Michael~McCourt, Smadar~Naoz, Ryan~O'Leary, Enrico~Ramirez-Ruiz, Lorenzo~Sironi, and Nicholas~Stone for thoughtful discussions. This work was supported in part by Einstein grant PF3-140108 (J.~G.). J.~G. would like to thank the Aspen Center for Physics (NSF Grant \#1066293) for their generous hospitality. First paper draft was originally completed circa 2018; paper was updated in 2026 with the assistance of the Anthropic AI models Claude Fable and Opus 5. All references and AI edits were hand-checked by the authors.\footnote{The manuscript source, the bibliography, and the scripts that solve the equilibrium system and generate the figures are available at \url{https://github.com/guillochon/tde-engine}.}
\end{sloppypar}

\bibliographystyle{apj}
\bibliography{engineNotes}

\end{document}